\documentclass[final,12pt]{elsarticle}

\usepackage{graphicx}

\usepackage{amssymb}
\usepackage{amsmath}
\biboptions{sort&compress}
\newcommand{\nc}{\newcommand}		
\nc{\vc}[1]	{\mbox{\boldmath $#1$}}	
\nc{\del}       {\partial}              
\nc{\bra}       {\langle}               
\nc{\ket}       {\rangle}               
\nc{\bras}[1]   {\langle #1|}           
\nc{\kets}[1]   {|#1\rangle}            
\nc{\mapleft}[1]{			
 \smash{\mathop{\,			%
  \hbox to 1.5cm{\rightarrowfill}\, }\limits_{#1}}}
\nc{\beq}     {\begin{eqnarray}}
\nc{\eeq}    {\end{eqnarray}}
\nc{\fra}     {\frac{1}{2}}
\nc{\nn}    {\\\nonumber}
\nc{\bx}     {\bold x}
\nc{\be}    {\bold e}
\nc{\bE}    {\bold E}
\nc{\bB}    {\bold B}
\nc{\bl}    {\bold l}
\nc{\bS}    {\bold S}
\nc{\bs}    {\bold s}
\nc{\bA}    {\bold A}
\newcommand{\f}[2]{\frac{d^#2 #1}{(2\pi)^#2}}
\newcommand{\tr}{\mathrm{tr}}
\newcommand{\mcl}{\mathcal}

\journal{Nuclear Physics A}

\begin{document}

\begin{frontmatter}



\title{Quark-Hadron Matter at Finite Temperature and Density in a Two-Color PNJL model}


\author[label1]{Shotaro Imai}
\author[label1]{Hiroshi Toki}
\author[label2,label3]{Wolfram Weise}

\address[label1]{Research Center for Nuclear Physics (RCNP), Osaka University, Ibaraki, Osaka 567-0047, Japan}

\address[label2]{
Physik-Department, Technische Universit\"{a}t M\"{u}nchen, D-85747 Garching, Germany}

\address[label3]{
ECT*, Villa Tambosi, I-38123 Villazzano (Trento), Italy}

\begin{abstract}
Quark-hadron matter at finite temperature and density is studied using a two-flavor, color $SU(2)$ (P)NJL model. The hadronic effective Lagrangian, derived by bosonization of the quark fields and renormalized using the Eguchi method, emerges in the form of an extended linear $\sigma$ model with meson and diquark-baryon fields.  Chiral and diquark condensates are studied as functions of temperature and baryon density.   Masses of mesons and diquark-baryons are calculated with and without the Polyakov loop effect.  We investigate the equation of state of quark-hadron matter by taking into account the contributions of mesons and diquark-baryons in addition to the quark quasiparticles.  

\end{abstract}


\end{frontmatter}


\section{Introduction}
\label{intro}

Quarks and gluons are basic ingredients of quantum chromodynamics (QCD).  At low temperature and density, the quarks and gluons are confined in colorless composites, the hadrons.  However, at high temperature and/or high density, the quarks and gluons are de-confined and become active degrees of freedom. Hence, for the discussion of strongly interacting matter in various situations, we ought to describe the dynamics of quarks and gluons and at the same time baryons and mesons in order to delineate the corresponding phases.

Lattice QCD (LQCD) simulations can handle strongly interacting matter at finite temperature and at very small baryon chemical potential.  As a consequence of the sign problem in LQCD, it is extremely difficult, however, to deal with the QCD quark-gluon matter at finite density covering broad ranges of real baryon chemical potential. The Nambu-Jona-Lasinio (NJL) model~\cite{Nambu1961,Vogl1991,Hatsuda1994,Klevansky1992} is often used as an alternative, schematic approach to strongly interacting matter~\cite{Blaschke2004, Bernard1988,Sadzikowski2001,Brauner2008,Zhuang1994}, based on chiral symmetry and its spontaneous breaking.
The NJL model is further generalized by introducing the Polyakov loop to account for important thermodynamical aspects of color confinement~\cite{Fukushima2003a, Fukushima2004, Ratti2006}. 
This PNJL model is by now widely used for the discussion of quark-hadron matter at finite temperature and density~\cite{Hansen2007, Roessner2008, Roessner2007, Fukushima2008, Megias2006}.
The PNJL model results can be compared directly with lattice QCD at finite temperature and zero baryon density.  There is, however, a basic conceptual problem when applying the PNJL model at finite baryon density: color-singlet baryon formation is not accounted for.  While color-nonsinglet degrees of freedom are suppressed in the ``baryonic'' phase by the Polyakov loop, three quarks  coupled to a color-singlet are still delocalized and spread over all space instead of being confined in localized baryonic clusters.

In this context it is instructive to study the simpler case of two-color QCD.  With the gauge group $SU(2)_{c}$, the corresponding lattice QCD approach does not have the sign problem, and LQCD simulations can be performed at any baryon chemical potential $\mu$.  Quarks carry baryon number $\fra$.  Baryons are diquarks, {\it i.e.} spin singlet or triplet bosons in this theory.  The resulting physics is qualitatively different from ``real'' QCD with $N_{c}=3$.  Nonetheless, exploring the $N_{c}=2$ theory and designing (P)NJL type models in which both mesons (quark-antiquark modes) and baryons (diquarks) emerge as active hadronic degrees of freedom, can teach important lessons about the thermodynamics and the phase structure of strongly interacting quark-hadron matter at non-zero baryon chemical potential.

Several LQCD studies for color $SU(2)$ are available~\cite{Hands2000, Hands2001, Hands2008, Kogut2001,Skullerud2004, Takahashi2010} and provide equations of state and hadron masses of two-color quark-hadron matter at both finite temperature and density.  It is then an interesting question to what extent these LQCD results can be understood and interpreted in terms of models, such as the PNJL approach, in order to identify leading mechanisms and basic symmetry breaking patterns.  Our work is aimed in that direction.  It presents major steps beyond a previous NJL model calculation~\cite{Ratti2004} performed in a similar context, and it complements other related two-color QCD studies~\cite{Kogut1999,Kogut2000, Lenaghan2002, Splittorff2001,Splittorff2002a, Harada2010, Andersen2010,  Brauner2009, Kanazawa2009, Nishida2004, Strodthoff2012, Zhang2010, He2010, Sun2007, Ebert2011}.



In this paper, we study the color $SU(2)$ NJL model using bosonization and renormalization techniques to construct the meson and diquark-baryon effective Lagrangian.  The derivation of the complete hadron Lagrangian follows the Eguchi method \cite{Eguchi1976} and renormalization lead to a generalized linear $\sigma$ model with Pauli-G\"{u}rsey symmetry.  The partition function and the thermodynamical potential are derived with and without incorporation of Polyakov loop effects. The equation of state is deduced and the phase diagram is studied in detail.  We investigate the excitation spectrum of quark-hadron matter at various temperatures and densities.  We then include the effect of hadrons in the thermodynamical potential and calculate the pressure and the quark density using the Gaussian approximation~\cite{Diener2008} of the hadron Lagrangian, taking only the mass terms into account.

This paper is organized as follows.  In Sect.$\,$2, the NJL model for the case of color $SU(2)$ with two flavors is introduced, bosonization is performed and an extended linear $\sigma$ model is derived.   The thermodynamical potential of the meson and diquark-baryon fields is deduced in Sect.$\,$3.  In Sect.$\,$4, we present numerical results and compare with LQCD results.  Sect.$\,$5 is devoted to conclusions and outlook.

\section{Two color NJL model}
\label{NJL}
\subsection{NJL Lagrangian and auxiliary fields}

We start with the NJL Lagrangian for color $SU(2)$ and two quark flavors: 
\beq
\mcl{L}_{NJL}&=&\bar{\psi}(i\gamma_\nu \partial^\nu-m_0+\gamma_0\mu)\psi+\frac{G_0}{2}[(\bar{\psi}\psi)^2+(\bar{\psi}i\gamma_5\vec{\tau}\psi)^2] \nn
&&+\frac{H_0}{2}(\bar{\psi}i\gamma_5t_2\tau_2C\bar{\psi}^T)(\psi^TCi\gamma_5t_2\tau_2\psi)~.
\eeq
The quark fields are $\psi=(u,d)^T$, with masses $m_u=m_d=m_0$.  The transpose of $\psi$ is denoted by $\psi^T$.  The $SU(2)$ isospin and color matrices are written $\tau_{i}$ and $t_{j}$, respectively.  We introduce the quark chemical potential $\mu$ for the study of finite baryon density, with the baryon chemical potential $\mu_{B}=2\mu$.  The diquark-baryon fields involve the charge conjugation operator for fermions, $C=i\gamma_0\gamma_2$, together with $t_{2}$ and $\tau_{2}$ to form color and isospin singlets.  
Starting from a color current-current interaction proportional to $(\bar \psi \gamma_{\mu}t_{j}\psi)(\bar \psi \gamma^{\mu}t_{j}\psi)$,
a Fierz transformation provides equal coupling constants $G_0$ for the meson channels and $H_0$ for the diquark channels, featuring the Pauli-G\"{u}rsey (PG) symmetry.   We keep, however, the coupling constants independent so that we are able to study as well cases in which PG symmetry is not exactly realized.

The bosonization technique is now used to write the Lagrangian in terms of auxiliary meson fields, $\sigma(x), \vec{\pi}(x)$, and diquark fields, $\Delta(x), \Delta^\ast(x)$~\cite{Eguchi1976,Kahana1990,Ebert1991,Ebert1998}.  The partition function of the NJL model including these auxiliary fields is
\beq
\label{partition}
 Z = \int \mcl{D} \bar{\psi} \mcl{D}\psi \mcl{D} \sigma \mcl{D} \vec\pi \,\mcl{D} \Delta \mcl{D} \Delta^* \exp \left(i\int d^4x \mcl{L}_{aux}\right)~,
\eeq
with the Lagrangian
\beq
\mcl{L}_{aux}&=&\fra(
\begin{array}{cc}
 \bar{\psi} & \psi^TC
\end{array})
\left(
\begin{array}{cc}
 S^{-1}(\mu)&g_d\gamma_5t_2\tau_2\Delta(x)\\
 -g_d\Delta^\ast(x) \gamma_5t_2\tau_2&S^{-1}(-\mu)
\end{array}\right)\left(
\begin{array}{c}
 \psi\\
 C\bar{\psi}^T
\end{array}\right)\nn&&
-\frac{1}{2}M_s^2(\sigma^2(x)+\vec{\pi}^{\,2}(x))-\frac{1}{2}M_d^2\Delta^\ast(x)\Delta(x)~,
\eeq
with $S^{-1}(\pm \mu)=i \gamma_\mu \partial^\mu -m_0 \pm \gamma_0 \mu-g_0(\sigma(x) \pm i\gamma_5\vec{\pi}(x)\cdot\vec{\tau})$.
We have introduced a meson coupling constant $g_0$ and a diquark coupling constant $g_d$ together with bare scalar and diquark masses, $M_s$ and $M_d$, in preparation of the standard renormalization scheme.  The meson and diquark coupling constants and masses are related as $\frac{g_0^2}{M_s^2}=G_0$ and $\frac{g_d^2}{M_d^2}=H_0$.  

Introducing a chiral order parameter $\sigma_{0}$ (proportional to the quark condensate $\bra \bar \psi \psi \ket$) as $\sigma(x)=\sigma_0+s(x)$ and the diquark condensates, $\Delta_{0}$ and $\Delta_{0}^{*}$, as $\Delta(x)=\Delta_0+d(x)$ and  $\Delta^\ast(x)=\Delta_0^\ast+d^\ast(x)$, and integrating out the quark fields, the effective Lagrangian entering the partition function (\ref{partition}) becomes
\beq
\label{effectivelagrangian}
\mcl{L}_{eff}&=& -\frac{i}{2}\tr( \ln\hat{S}^{-1}+\ln(1+\hat{S}\hat{K}) ) \nn &&-\frac{1}{2}M_s^2 \sigma_0^2- \frac{1}{2} M_s^2(s^2(x)+\vec{\pi}^2(x))-M_s^2\sigma_0s(x)\nn&&
-\frac{1}{2}M_d^2\Delta_0^\ast\Delta_0
-\frac{1}{2}M_d^2(\Delta_0^\ast d(x)+d^\ast(x)\Delta_0)-\frac{1}{2}M_d^2d^\ast(x) d(x)~.
\eeq
The trace is taken over spin, flavor and color spaces.
The matrices $\hat S^{-1}$ and $\hat K$ are defined as,
\beq
 \hat{S}^{-1}&=&
\left( \begin{array}{cc}
 S_{0}^{-1}(\mu)&\Delta^{-}\\
 \Delta^{+}&S^{-1}_{0}(-\mu)
\end{array}\right)~,
\eeq
\beq
\hat{K}&=&
\left( \begin{array}{cc}
 -g_{0}(s(x)+i\gamma_5\vec{\tau}\cdot\vec{\pi}(x))&g_{d}\gamma_5t_2\tau_2d(x)\\
 -g_{d}d^{\ast}(x)\gamma_5t_2\tau_2&-g_{0}(s(x)-i\gamma_5\vec{\tau}\cdot\vec{\pi}(x))
\end{array}\right)~,
\eeq
with $\Delta^{-}=g_{d}\gamma_5t_2\tau_2\Delta_{0}, \Delta^{+}=-g_{d}\Delta_{0}^{\ast}\gamma_5t_2\tau_2$ and $S_0^{-1}(\pm \mu)=i \gamma_\mu \partial^\mu -m \pm \gamma_0 \mu$.   A dynamical quark mass is defined as $m=m_0+g_0\sigma_0$.
The Nambu-Gorkov quark propagator matrix $\hat{S}$ is determined by solving $\hat{S}\hat{S}^{-1}=1$ and expressed as:
\beq
 \hat{S}=
\left( \begin{array}{cc}
 G^+&H^-\\
 H^+&G^-
\end{array}\right),
\eeq
with the components
\beq
 G^{\pm}&=&(S_0^{-1}(\pm\mu)-\Sigma^{\pm})^{-1}\nn
 \Sigma^{\pm}&=&\Delta^{\mp}S_{0}(\mp\mu)\Delta^{\pm}\nn
 H^{\pm}&=&-S_0(\mp\mu)\Delta^{\pm}G^{\pm}.
\eeq

A simple form for the components of the Nambu-Gorkov propagator is found introducing the energy projectors onto states of positive and negative energy for free massive spin $1/2$ quasi-particles following Huang {\it et al.} \cite{Huang2005,Huang2002},
\beq
 \Lambda_{\pm}=\frac{1}{2}\left(1\pm\frac{\gamma_0(\vec{\gamma}\cdot\vec{p}+m)}{E_p}\right),
\quad
\tilde{\Lambda}_{\pm}=\frac{1}{2}\left(1\pm\frac{\gamma_0(\vec{\gamma}\cdot\vec{p}-m)}{E_p}\right),
\eeq
with $E_p=\sqrt{\vec{p}^{\,2}+m^2}$.
These operators satisfy the projection properties
$ \Lambda_{\pm}\Lambda_{\pm}=\Lambda_{\pm}$, $\Lambda_{\pm}\Lambda_{\mp}=0$, $\Lambda_{+}+\Lambda_{-}=1$, and similar relations for $\tilde{\Lambda}$.  Furthermore, the two projection operators are related as
$\gamma_0\Lambda_{\pm}\gamma_0=\tilde{\Lambda}_{\mp}$ and $\gamma_5\Lambda_{\pm}\gamma_5=\tilde{\Lambda}_{\pm}$.
The Dirac propagator can be written
\beq
\label{prop}
S_0(\pm\mu)=\frac{\Lambda_{+}\gamma_0}{p_0-E_p^{\mp}}+\frac{\Lambda_{-}\gamma_0}{p_0+E_p^{\pm}}~.
\eeq

Taking its inverse we arrive at
\beq
\label{propi}
S_0^{-1}(\pm\mu)=(p_0-E_p^{\mp})\gamma_0\Lambda_{+}+(p_0+E_p^{\pm})\gamma_0\Lambda_{-}~,
\eeq
with $E_p^{\pm}=E_p\pm\mu$.
It follows that the components of the Nambu-Gorkov propagator are written as
\beq
 G^{\pm} &=&\frac{p_0+ E_p^{\mp}}{p_0^2-(E_\Delta^{\mp})^{2}}\Lambda_{+}\gamma_0 +\frac{p_0- E_p^{\pm}}{p_0^2-(E_\Delta^{\pm})^{2}}\Lambda_{-} \gamma_0~, \\
 H^{\pm}&=&\frac{\Delta^{\pm}}{p_0^2-(E_\Delta^{\pm})^{2}}\tilde{\Lambda}_+
+\frac{\Delta^{\pm}}{p_0^2-(E_\Delta^{\mp})^{2}}\tilde{\Lambda}_-~.
\eeq
The quasi-particle energy is $E^{\pm}_{\Delta}=\sqrt{(E_p^{\pm})^{2}+g_{d}^{2}|\Delta_{0}|^2}$ with the dynamical quark mass $m=m_0+g_0\sigma_0$ and the diquark gap $\Delta_0$. 

\subsection{Mean field approximation}

Keeping only the expectation values $\bra \sigma(x)\ket=\sigma_{0}$, $\bra \Delta(x)\ket=\Delta_{0}$ and $\bra \Delta^{*}(x)\ket=\Delta_{0}^{*}$, {\it i.e.} dropping  the fluctuating meson and diquark fields $s(x)$, $\vec\pi(x)$, $d(x)$ and $d^{*}(x)$, one arrives at the mean-field Lagrangian
\beq
\mcl{L}_{MF}= -\frac{i}{2} \tr \ln \hat{S}^{-1}-\frac{1}{2}M_s^2\sigma_0^2-\frac{1}{2}M_d^2\Delta^{*}_{0}\Delta_{0}~.
\eeq
The first term is calculated using the relation $\tr\ln \hat{S}^{-1}=\ln \det \hat{S}^{-1}$. The determinant is taken over spin, flavor, color and momentum space and calculated using Eq. \eqref{propi} together with the properties of the projection operators \cite{Huang2005, Huang2002}.  One finds
 \beq
  \det\hat{S}^{-1}=\sqrt{(p_0^2-(E_\Delta^{-})^{2})(p_0^2-(E_\Delta^{+})^{2})} ~.
 \eeq
Temperature is now introduced using the Matsubara formalism,
\beq
i\int \frac{d^4 p}{(2\pi)^4} \rightarrow -T\sum_n \int \frac{d^3 p}{(2\pi)^3}~,
\eeq
with the replacement $p_0\rightarrow i\omega_n$ where $\omega_n=(2n+1)\pi T$ are the fermionic Matsubara frequencies.
Taking the frequency sum, the thermodynamical potential becomes
\beq
\label{omegamean1}
\Omega_{MF} &=& -\frac{T}{V}\ln Z=- \tr \int \frac{d^3 p}{(2\pi)^3} (E_\Delta^++E_\Delta^-)\nn&&-2\, \tr \int 
\frac{d^3 p}{(2\pi)^3} T[ \ln(1+e^{-\beta E_\Delta^+}) + \ln(1+e^{-\beta E_\Delta^-})]\nn&&+\frac{1}{2}M_s^2\sigma_0^2+\fra M_d^2 |\Delta_0|^2~.
\eeq
Here we have written $\Delta^{*}_{0}\Delta_{0}=|\Delta_{0}|^{2}$ and the inverse of the temperature $T$ as $\beta=1/T$.

The derivatives of the thermodynamical potential (\ref{omegamean1}) with respect to $\sigma_{0}$ and $|\Delta_0|$ determine the chiral condensate and the diquark condensate at the minimum of $\Omega_{MF}$:
\beq
\label{gap}
\nonumber
\frac{\partial \Omega_{MF}}{\partial \sigma_{0}}&=&- \tr\int\f{p}{3}\frac{g_{0}m}{E_p}\left[\frac{E_p^+}{E_\Delta^+}\left(1-2f(E_\Delta^+)\right)+\frac{E_p^-}{E_\Delta^-}\left(1-2f(E_\Delta^-)\right)\right]\\&&
+M_{s}^{2}\sigma_{0}=0~,\\
\label{gapdelta}
\nonumber
\frac{\partial \Omega_{MF}}{\partial |\Delta_0|}&=&-|\Delta_0| \tr\int\f{p}{3}\left[\frac{1}{E_\Delta^+}\left(1-2f(E_\Delta^+)\right)+\frac{1}{E_\Delta^-}\left(1-2f(E_\Delta^-)\right)\right]\\&&
+M_d^2|\Delta_0|=0~,
\eeq
with the Fermi distribution function $f(E)=(1+e^{\beta E})^{-1}$.  The chemical potential $\mu$ is included in the definition of $E_{\Delta}^{\pm}$.

\subsection{Hadron Lagrangian from bosonization}
\label{effect}

We would like to extract the properties of mesons and diquark-baryons at finite temperature and density.  For this purpose, we expand the logarithmic term in Eq. \eqref{effectivelagrangian} as
\beq
-\frac{i}{2}\tr\ln(1+\hat{S}\hat{K})=-\frac{i}{2}\tr\sum_{k=1}^{\infty}\frac{(-1)^{k+1}}{k}(\hat{S}\hat{K})^k
\equiv\sum_{k=1}^{\infty}U^{(k)}~,
\eeq
where
\beq
\label{expansion}
 U^{(k)}=-\frac{i(-1)^{k+1}}{2k}\tr(\hat{S}\hat{K})^k~.
\eeq
The matrix
\beq
\hat{S}\hat{K}=
\left(\begin{array}{cc}
A & B\\
C & D
\end{array}\right)~,
\eeq
has elements
\beq
A&=& -G^+g_0\,(s(x)+i\gamma_5\vec{\tau}\cdot\vec{\pi}(x))-H^-g_d\,d^*(x) \gamma_5 t_2 \tau_2~,\nn
B&=& G^+g_d\,\gamma_5t_2\tau_2 d(x)- H^- g_0\,(s(x)-i\gamma_5\vec{\tau}\cdot\vec{\pi}(x))~,\nn
C&=& -H^+g_0\,(s(x)+i\gamma_5\vec{\tau}\cdot\vec{\pi}(x))-G^-g_d\,d^\ast(x)\gamma_5t_2\tau_2~,\nn
D&=& H^+g_d\,\gamma_5t_2\tau_2d(x)-G^-g_0\,(s(x)-i\gamma_5\vec{\tau}\cdot\vec{\pi}(x))~.
\eeq 
With these expressions we are able to derive the Lagrangian involving the meson and diquark-baryon fields.

\subsection{The gap equations (k=1)}

We first work out the case of $k=1$, that is $U^{(1)}=-\frac{i}{2}\tr\hat{S}\hat{K}$, to derive the mass gap equations, which correspond to the minimization condition \eqref{gap} of the thermodynamical potential in the mean-field approximation described in the previous subsection.
\beq
\nonumber
U^{(1)}&=&-\frac{i}{2}\tr\int d^4x\f{p}{4} \left[g_0(-G^{+}-G^{-})s(x)+g_0 (-G^{+}i\gamma_5+G^{-}i\gamma_5)\vec{\tau}\cdot\vec{\pi}(x)\right.\nn&&\left.-g_dH^{-}\gamma_5t_2\tau_2d^\ast(x)+g_dH^{+}\gamma_5t_2\tau_2d(x)\right]\\
&=&\int d^4x\Bigl(\Gamma_s \, s(x)+\Gamma_d\, d(x)+\Gamma_{d^\ast} \, d^\ast(x)\Bigr)~.
\eeq
The trace of the Dirac matrix in the pion term gives zero and we have dropped this term here.
The result for $\Gamma_s$ is
\beq
\label{gs}
 \Gamma_{s}&=& \frac{i}{2}g_0\,\tr\int\f{p}{4}(G^{+} + G^{-})=2img_0\,\tr_{fc}\int\f{p}{4}\times  \nn&& \left[\frac{1}{p_0^2-(E_\Delta^{+})^{2}}+\frac{1}{p_0^2-(E_\Delta^{-})^{2}}+\frac{\mu}{E_p}\left(\frac{1}{p_0^2-(E_\Delta^{+})^{2}}-\frac{1}{p_0^2-(E_\Delta^{-})^{2}}\right)\right]~.
\eeq
The trace $\tr$ is taken over spin, flavor and color spaces  and $\tr_{fc}$ is over flavor and color only.  This expression agrees with the mean-field equation (\ref{gap}) at $T=0$, when explicitly writing $E_{p}^{\pm}=E_{p}\pm\mu$, so that
\beq
\label{k1}
\frac{\partial \Omega_{MF}}{\partial \sigma_{0}}&=&-g_{0}\,m\,\tr\int\f{p}{3}\left[\frac{1}{E_{\Delta}^{+}}\left(1-2f(E_{\Delta}^+)\right)+\frac{1}{E_{\Delta}^{-}}\left(1-2f(E_{\Delta}^-)\right)\right.\nn
&&\left.+\frac{\mu}{E_p}\left(\frac{1}{E_{\Delta}^{+}}\left(1-2f(E_{\Delta}^+)\right)-\frac{1}{E_{\Delta}^{-}}\left(1-2f(E_{\Delta}^-)\right)\right)\right]+M_{s}^{2}\sigma_{0}
=0 ~.
\eeq

The momentum space integrals (\ref{gs}) are divergent and must be treated accordingly.  For the second term under the integral we write:
\beq
&&\frac{\mu}{E_p}\left(\frac{1}{p_0^2-(E_\Delta^{+})^{2}}-\frac{1}{p_0^2-(E_\Delta^{-})^{2}}\right)\nn&=&
\frac{\mu}{E_p}\left(\frac{1}{p_0^2-(E_{\Delta}^{-})^{2}-4\mu E_p}-\frac{1}{p_0^2-(E_{\Delta}^{+})^{2}+4\mu E_p}\right)\nn
&\sim &\frac{\mu}{E_p}\left[\frac{1}{p_0^2-(E_{\Delta}^{-})^{2}}\left(1+\frac{4\mu E_p}{p_0^2-(E_{\Delta}^{-})^{2}}\right)-\frac{1}{p_0^2-(E_{\Delta}^{+})^{2}}\left(1-\frac{4\mu E_p}{p_0^2-(E_{\Delta}^{+})^{2}}\right)\right]\nn
&=&2\mu^2\left(\frac{1}{(p_0^2-(E_\Delta^{+})^{2})^2}+\frac{1}{(p_0^2-(E_\Delta^{-})^{2})^2}\right)~,
\eeq
where in the third line only those terms have been kept that are divergent when taking the momentum integral.  Using this expression $\Gamma_s$ can be written as
\beq
 \Gamma_s = 2mg_0(I_2-2\mu^2I_0)~,
\eeq
with the divergent integrals $I_2$ and $I_0$ defined as follows:
\beq
 I_2&=&i\tr_{fc}\int\f{p}{4}\left(\frac{1}{p_0^2-(E_\Delta^{+})^{2}}+\frac{1}{p_0^2-(E_\Delta^{-})^{2}}\right)~,\\
 I_0&=&-i \tr_{fc} \int\f{p}{4}\left(\frac{1}{(p_0^2-(E_\Delta^{+})^{2})^2}+\frac{1}{(p_0^2-(E_\Delta^{-})^{2})^2}\right)~.
\eeq

The result for $\Gamma_d$ is
\beq
\Gamma_{d} =-\frac{i}{2}g_d\, \tr \int\f{p}{4}H^+\gamma_5 t_2 \tau_2 = g_d^2\,\Delta_0^\ast I_2~,
\eeq
while
\beq
\Gamma_{d^*}=g_d^2\,\Delta_0 I_2~.
\eeq
At this stage the effective Lagrangian is $\mcl{L}_{eff}=\mcl{L}_{MF}+\mcl{L}^{(1)}$ with
\beq
\mcl{L}^{(1)}=2g_0 m(I_2-2\mu^2I_0)s+g_d^2 I_2(\Delta_0^\ast  d+\Delta_0 d^{*})~.
\eeq
It involves the leading-order couplings of the condensates to the meson and diquark fluctuations, $s(x)$, $d(x)$ and $d^{*}(x)$.

$\Gamma_s$ together with other terms linear in $s$ in the effective Lagrangian gives the mass gap equation by requiring
\beq
 -M_{s}^2\sigma_0+2g_0 m(I_2-2\mu^2I_0)=0~.\label{mass}
\eeq
We recall that the relation between $m$ and $\sigma_{0}$ is $m=g_{0}\sigma_{0}+m_{0}$.
With the $\Gamma_d$ and $\Gamma_{d^*}$ terms and the linear diquark-baryon terms in the Lagrangian, we obtain the diquark condensate equation,
\beq
 -M_d^2 \Delta_0 +2 g_d^2\,\Delta_0 I_2=0~.
\eeq

For later use, we write $I_2$ and $I_0$ as functions of temperature and chemical potential using the Matsubara formalism:
\beq
I_2&=&- \tr_{fc} T\sum_n \int \frac{d^3p}{(2\pi)^3} \left(\frac{1}{(i\omega_n)^2-(E_{\Delta}^{+})^{2}}+\frac{1}{(i\omega_n)^2-(E_\Delta^{-})^{2}}\right)\nn
&=& \tr_{fc} \int \frac{d^3p}{(2\pi)^3} \left( \frac{1}{2E_\Delta^+}(1-2f(E_\Delta^+))+\frac{1}{2E_\Delta^-}(1-2f(E_\Delta^-))\right)~.\label{i2}
\eeq
Proceeding analogously in Eq. \eqref{gs}, the mass gap equation  \eqref{mass} is modified accordingly and agrees with Eq. \eqref{k1}.

Next, we work out $I_0$:
\beq
 I_0&=&\tr_{fc}T\sum_{n}\int\f{p}{3}\left(\frac{1}{((i\omega_n)^2-(E_\Delta^{+})^{2})^2}+\frac{1}{((i\omega_n)^2-(E_\Delta^{-})^{2})^2}\right)\nn
&=&\tr\int\f{p}{3}\left[\frac{1}{4(E_\Delta^{+})^{3}}\left(1-2f(E_\Delta^+)-2E_\Delta^+\,\beta\, e^{\beta E_\Delta^+}f^2(E_\Delta^+)\right)\right.\nn&&\left.
+\frac{1}{4(E_\Delta^{-})^{3}}\left(1-2f(E_\Delta^-)-2E_\Delta^-\,\beta\, e^{\beta E_\Delta^-}f^2(E_\Delta^-)\right)\right]~.
\eeq
With these expressions, the gap equations are - as expected - identical to the expressions found by taking derivatives of $\Omega_{MF}$ with respect to the condensates (see Eqs. (\ref{gap}) and (\ref{gapdelta})).  The divergent integrals $I_{0}$ and $I_{2}$ are calculated introducing a momentum cut-off $\Lambda$.

\subsection{Mass and kinetic energy terms (k=2)}

The kinetic energies and the mass terms of mesons and diquark-baryons emerge from Eq. (\ref{expansion}) at the order of $k=2$:
\beq
\label{k2}
 U^{(2)}=\frac{i}{4}\tr(\hat{S}\hat{K})^2 = \int d^4x \, d^4y \sum_{ij}\Gamma_{ij}(x-y)\phi_{i}(x)\phi_{j}(y)~,
\eeq
where $\phi_{i}=\sigma, \vec{\pi}, d, d^{*}$.
The non-local propagator terms $\Gamma_{ij}$ are expanded as
\beq
 \Gamma(x-y)&=&\int \f{q}{4}e^{iq(x-y)}\Gamma(q)\nn
&\simeq& \delta(x-y)\Gamma(0)-i\partial^{\mu}\delta(x-y)\partial_{\mu}\Gamma(q)\mid_{q=0}\nn&&
-\frac{1}{2}\partial^{\mu}\partial^{\nu}\delta(x-y)\partial_{\mu}\partial_{\nu}\Gamma(q)\mid_{q=0}~.
\eeq

Detailed results for the quantity $\Gamma(q)\mid_{q=0}$ and the derivative terms $\partial_{\mu}\Gamma(q)\mid_{q=0}$ and $\partial_{\mu}\partial_{\nu}\Gamma(q)\mid_{q=0}$ are given in Appendix A.  The emerging piece of the effective Lagrangian at order $k=2$ is:
\beq
\label{second}
\mcl{L}^{(2)}&=& \fra g_0^2 I_0 (\partial_\mu \vec{\pi})^2+ g_0^2( I_2 -2 I_0 \mu^2) \vec{\pi}^2\nn
&+& \frac{1}{2}g_0^2I_0 (\partial_\mu s)^2+g_0^2(I_2-2I_0(\mu^2+m^2) )s^2 \nn
&+& \frac{1}{2}g_d^2I_0(\partial^\mu d^{*})(\partial_\mu d)+g_d^2(I_2- I_0g_{d}^{2}|\Delta_{0}|^2)d^{*}d\nn
&-&ig_d^2I_0\mu(d^*\partial_0d-d\partial_0d^*)-\frac{1}{2}g_d^4I_0(\Delta^{*2}_{0}d^2+\Delta^{2}_{0}d^{*2})\nn
&-&2g_0g_d^2I_0m\, s(\Delta_0^*d+\Delta_0d^*)~.
\eeq
The Lagrangian density has mass dimension 4, and $I_{0}$ and $I_{2}$ have mass dimensions 0 and 2, respectively.  Hence when $I_{0}$ appears in the coefficients there are factors of mass dimension 2 such as squares of the derivative $\partial_{\mu}$, the chemical potential $\mu$, the quark mass $m$ and the gap energy $\Delta_{0}$ and $\Delta_{0}^{*}$.  The pion term has a simple form due to the pseudo-scalar nature of the pion field, while the scalar boson $s(x)$ couples to the diquark-baryon fields through the diquark condensates.

\subsection{The coupling terms (k=3)}
The coupling vertices involving combinations of three mesons or diquark fields are obtained from the third-order term:
\beq
 U^{(3)}=-\frac{i}{6}\tr(\hat{S}\hat{K})^3&=&-\frac{i}{6}\tr(A^3+ABC+BCA+CAB\nn
&&+CBD+BDC+DCB+D^3)~.
\eeq
Non-local terms with derivatives of hadron fields vanish, and $U^{(3)}$ is constructed with only local terms.   The details of the derivation are found in Appendix A.  The final result for the $k=3$ part of the effective Lagrangian is:  
\beq
\label{third}
\mcl{L}^{(3)}&=&-2 g_0 I_0 m \left[g_{0}^{2}(s^2+\vec{\pi}^2)s+g_d^2 d^{*}d s\right]\nn
&&-g_d^2 I_0\left[g_0^2(s^2+\vec{\pi}^2)(\Delta_0 d^{*}+\Delta^*_0 d)+ g_d^2 d^{*}d(\Delta_0 d^*+\Delta_{0}^{*}d)\right]~.
\eeq
The coupling terms have mass dimension 3 due to the presence of altogether three meson and diquark-baryon fields.   Hence, the coefficients of all $k=3$ vertices have mass dimension one with factors such as $m, \Delta_{0}$ and $\Delta_{0}^*$, together with the dimensionless integral $I_{0}$.   Those terms with $m$ in the coefficient conserve the baryon number, while terms with $\Delta_{0}$ or $\Delta_{0}^{*}$ are baryon number changing.

\subsection{The interaction terms (k=4)}

Four-point interaction terms involving mesons and diquarks derive  from the fourth-order term,
\beq
 U^{(4)}=\frac{i}{8}\tr(\hat{S}\hat{K})^4~.
\eeq
The non-local pieces vanish. Details of the derivation are again given in Appendix A.  The final expression for the $k=4$ part of the effective Lagrangian is:
\beq
\label{fourth}
\mcl{L}^{(4)}&=&-g_0^4I_0s^2\vec{\pi}^2-g_0^2g_d^2I_0(s^2+\vec{\pi}^2)d^*d
 \nn &&
-\frac{1}{2}g_0^4I_0(s^4+\vec{\pi}^4)-\frac{1}{2}g_d^4I_0d^{*2}d^2\nn
&=&-\frac{1}{2} I_0\left[g_0^2 (s^2+\vec{\pi}^2)+g_d^2d^*d\right]^2~.
\eeq
It generates interactions between diquarks, scalar mesons and pions in all possible combinations.

\subsection{Identification with a generalized linear $\sigma$ model}

The limiting case of exact Pauli-G\"{u}rsey symmetry in color $SU(2)$ is realized with $g_{0}=g_{d}$ and $M_{s}=M_{d}$.  The fourth order Lagrangian density is written as
\beq
\mcl{L}^{(4)}=-\frac{1}{2} g_0^4 I_0( s^2+\vec{\pi}^2+d^*d)^2~.
\eeq
This form and the structure of the other terms suggest, not surprisingly, that the hadron Lagrangian is related to a generalized linear $\sigma$ model.  To demonstrate this, we introduce a fourth-order extended $\sigma$ model Lagrangian density with $\sigma$, $\vec{\pi}$, $\Delta$ and $\Delta^{*}$ fields:
\beq
\label{sigmafourth}
\mcl{L}_{\sigma}^{(4)}=-\frac{1}{2} g_0^4 I_0 ( \sigma^2+\vec{\pi}^2+\Delta^*\Delta)^2~.
\eeq
Separating again mean fields and fluctuations by the relations $\sigma=\sigma_{0}+s(x)$ and $\Delta=\Delta_{0}+d(x)$ together with its complex conjugate,  one finds
\beq
\sigma^{2}+\vec{\pi}^{2}+\Delta^{*}\Delta&=&(\sigma_{0}+s(x))^2+\vec{\pi}^2(x)+(\Delta_{0}+d(x))(\Delta_{0}^*+d^{*}(x))\nn&=&\sigma_{0}^{2}+\Delta_{0}\Delta_{0}^{*}+2\sigma_{0}s(x)+\Delta_{0}d^{*}(x)+\Delta_{0}^{*}d(x)\nn
&&+s^{2}(x)+\vec{\pi}^{2}(x)+d^{*}(x)d(x)~,
\eeq
and the 4-th order Lagrangian (\ref{sigmafourth}) is rewritten as:
\beq
\mcl{L}_{\sigma}^{(4)}&=&-\frac{1}{2} g_0^4 I_0[v_{0}^4+2v_{0}^2(2\sigma_{0}s+\Delta_{0}d^*+\Delta_{0}^{*}d)\nn
&&+2v_{0}^2(s^2+\vec{\pi}^2+d^{*}d)+4\sigma_{0}^2s^2+2|\Delta_{0}|^2d^{*}d\nn
&&+\Delta_{0}^2d^{*2}+\Delta_{0}^{*2}d^2+4\sigma_{0}\Delta_{0}sd^{*}+4\sigma_{0}\Delta_{0}^{*}sd\nn
&&+4\sigma_{0}(s^3+sd^{*}d+s\vec{\pi}^2)+2\Delta_{0}(d^{*2}d+s^2d^{*}+\vec{\pi}^{2}d^{*})\nn&&+2\Delta_{0}^{*}(d^{*}d^{2}+s^2d+\vec{\pi}^{2}d)+(s^2+\vec{\pi}^2+d^{*}d)^2]~,
\eeq
where we have defined $\sigma_{0}^2+|\Delta_{0}|^2=v_{0}^2$. 

Hence one can write the hadron Lagrangian in the compact form of a generalized linear $\sigma$ model with inclusion of diquark fields:
\beq
\label{unresigma}
\mcl{L}_{E\sigma}&=& \fra g_0^2 I_0 \left[(\partial_\nu \vec \pi)^2+ (\partial_\nu \sigma)^2+|(\partial_{\nu} -2i\mu\delta_{\nu 0})\Delta|^{2}\right]\nn&&-\fra (M_{s}^2 -2g_0^2 I_2 + 4g_0^2 I_0 \mu^2)(\sigma^{2}+ \vec \pi^2+\Delta^{*}\Delta)\nn
&&-\frac{1}{2} g_0^4 I_0\left[ ( \sigma^2+\vec \pi^2+\Delta^*\Delta)-v_{0}^{2}\right]^2~.
\eeq
In order to write the Lagrangian in this form, we have added $-2g_0^2 I_0 \mu^2\Delta^{*}\Delta$ to the mass term and combine with the $\sigma^{2}+\vec \pi^{2}$ terms to arrive at the compact symmetric mass term shown in the second line.  We then add the counter part, $+2g_0^2 I_0 \mu^2\Delta^{*}\Delta$, to the kinetic energy term proportional to $(\partial_{\nu}\Delta^{*})(\partial^{\nu}\Delta)$ and arrive at the compact form $\fra g_0^2 I_0|(\partial_{\nu} -2i\mu \delta_{\nu 0})\Delta|^{2}$. 

Consider next the explicit chiral symmetry breaking term.  Obviously, we have a term linear in the scalar field, $2g_0m_{0}(I_2-2\mu^2I_0)s$, which is proportional to the bare quark mass.  This suggests the explicit chiral symmetry breaking term of the form
\beq
\mcl{L}_{SB}=2g_0m_{0}(I_2-2\mu^2 I_0) \sigma~.
\eeq
This term provides a new mean-field mass equation including explicit chiral symmetry breaking,
\beq
-M_{s}^{2}\sigma_{0}+2g_{0}(m_{0}+g_{0}\sigma_{0})(I_{2}-2\mu^{2}I_{0})=0~.
\eeq
Self-consistent solution of this gap equation determines the dynamical quark mass, $m=m_{0}+g_{0}\sigma_{0}$, that appears also in the integrals $I_{2}$ and $I_{0}$. 

\subsection{Renormalization}

The hadron Lagrangian (\ref{unresigma}) extracted from the NJL model involves the divergent quark loop integrals $I_{2}$ and $I_{0}$.  The NJL approach is valid for quark momenta below a characteristic scale, $\Lambda \sim 0.6$ GeV, at which the integrals are cut off in practice.  Nonetheless, when written in the form (\ref{unresigma}) as a generalized linear sigma model, renormalization has to be performed.  We follow the Eguchi method~\cite{Eguchi1976}.  The kinetic term implies the wave function renormalization,
\beq
g_{0}^{2}I_{0}&=&Z_{M}^{-1}\nn
\sigma&=&Z_{M}^{\fra}\sigma_{R}\quad \vec \pi=Z_{M}^{\fra}\vec \pi_{R}\nn\Delta&=&Z_{M}^{\fra}\Delta_{R}\quad \Delta^{*}=Z_{M}^{\fra}\Delta_{R}^{*}~.
\eeq
The mass renormalization condition is
\beq
M_{s}^2 -2g_0^2 I_2 + 4g_0^2 I_0 \mu^2=0~.
\eeq
The coupling constant renormalization is written as
\beq
\label{lambda}
2g_{0}^{4}I_{0}&=&Z_{\lambda}^{-1}\lambda_{0}~,\nn
\lambda&=&Z_{M}^{2}Z_{\lambda}^{-1}\lambda_{0}~.
\eeq
The mean field is renormalized as
\beq
v=Z_{M}^{\fra}v_{0}~,
\eeq
and the explicit chiral symmetry breaking term is renormalized as
\beq
\label{barequark}
2g_0 m_{0}(I_2-2\mu^2I_0)&=&Z_{SB}^{-1}\varepsilon_{0}\nn
\varepsilon&=&Z_{M}^{\fra}Z_{SB}^{-1}\varepsilon_{0}~.
\eeq

When expressed in terms of renormalized fields and couplings, the effective Lagrangian (\ref{unresigma}) reads:
\beq
\label{resigma}
\mcl{L}_{E\sigma SB}&=& \fra \left[(\partial_\nu \vec \pi_{R})^2+ (\partial_\nu \sigma_{R})^2+|(\partial_{\nu} -2i\mu\delta_{\nu 0})\Delta_{R}|^{2}\right]\nn
&&-\frac{\lambda}{4}  ( \sigma_{R}^2+\vec \pi_{R}^2+\Delta_{R}^*\Delta_{R}-v^{2})^2+\varepsilon \sigma_{R}~.
\eeq
Together with the quark term, we have now a consistent Lagrangian to derive the thermodynamical potential for the interacting quark-hadron system.  The hadron dynamics itself is governed by the generalized linear $\sigma$ model Lagrangian (\ref{resigma}).  The explicit chiral symmetry breaking term, $\varepsilon \sigma_{R}$, is related to the bare quark mass $m_{0}$.  When the PG symmetry is not satisfied, the meson part and the diquark-baryon part are renormalized independently.

\section{Thermodynamics of mesons and diquark-baryons \\in the Gaussian approximation}

We now discuss the contributions of mesons and diquark-baryon fields to the thermodynamical potential, using the Gaussian approximation.  The starting point is the extended linear $\sigma$ model Lagrangian derived in the previous section.  The whole Lagrangian is written as
\beq
\mcl{L}=\mcl{L}_{MF}+\mcl{L}_{hadron}~.
\eeq
The mean field part has been worked out previously to provide the thermodynamical potential $\Omega_{MF}$.  The extended linear $\sigma$ model includes hadron fields up to fourth order.  The complete integrations over the hadron fields cannot be performed to this order.  In the present work we restrict ourselves to the Gaussian approximation \cite{Diener2008}, taking only the second order (mass) terms into account and perform the hadron integrals for the partition function as
\beq
Z=Z_q\int\mcl{D}s\mcl{D}\vec{\pi}\mcl{D}d\mcl{D}d^*\exp\left(i\int d^4x \mcl{L}_{hadron}^{(2)} \right)~,
\eeq
with
\beq
Z_q&=&\exp\left(-\frac{V}{T}\Omega_{MF}\right),\nn
\mcl{L}_{hadron}^{(2)}&=&\fra(\partial_\mu \vec \pi)^2-\fra m_{\pi}^2\vec \pi^2
+\frac{1}{2} (\partial_\mu s)^2-\fra m_s^2 s^2 \nn
&&+\frac{1}{2}(\partial^\mu d^{*})(\partial_\mu d)-\fra m_d^2 d^*d-i\mu(d^*\partial_0d-d\partial_0d^*)\nn
&&-\frac{1}{2}\Delta^{*2}d^2-\frac{1}{2}\Delta^{2}d^{*2}-2m\Delta^*sd-2m\Delta sd^*~.
\eeq
Here we have used the renormalized fields but omitted the index $R$ for simplicity.  For further convenience, we have introduced $\Delta=g_{d}\Delta_{0}$ and $\Delta^{*}=g_{d}\Delta_{0}^{*}$ in the above equation.  The renormalized masses are defined as
\beq
\label{renomass}
m_{\pi}^2&=&\left[M_{s}^2-2g_0^2(I_2-2 I_0 \mu^2)\right]/(g_{0}^{2}I_0)~,\nn
m_{s}^{2}&=&\left[M_{s}^2-2g_0^2(I_2-2 I_0(\mu^2+m^2)\right]/(g_{0}^{2}I_0)~,\nn
m_{d}^{2}&=&\left[M_{d}^2-2g_d^2(I_2-I_0|\Delta|^2)\right]/(g_{0}^{2}I_0)~.
\eeq
The meson and diquark-baryon masses (\ref{renomass}) found by bosonization and renormalization are identical to the masses obtained by solving the corresponding Bethe-Salpeter equations taking only the divergent integrals $I_{0}$ and $I_{2}$ in the quark-loop integrals.  Of course, using the Gaussian approximation means ignoring important interaction terms generated by $\mcl{L}^{(3)}$ and $\mcl{L}^{(4)}$.  These will be dealt with in separate forthcoming work.  A discussion of some qualitative features of such terms will follow the concluding section.

The hadronic partition function is
\beq
Z_{hadron}&=&\int\mcl{D}\Phi\mcl{D}\vec{\pi}\exp\left(-S^{(2)}\right),\nn
S^{(2)}&=&\fra\sum_{Q}\int d^3 x\left[\Phi^{\dagger}(x)M_n(Q)\Phi(x)+\vec{\pi}^{\dagger}(x)N_n(Q)\vec{\pi}(x)\right] ~,
\eeq
where $Q=(i\omega_n,\vec{q})$, with $\omega_n=2\pi n T$ being the boson Matsubara frequency and $\sum_{Q}=T\sum_n\int\f{q}{3}$.  The fields $\Phi$ are defined as $\Phi^{T}=(s, d, d^*)$ and $\vec{\pi}^{T}=(\pi^{+}, \pi^0, \pi^{-})$.  The matrix $M_n(Q)$ is the mass matrix in Euclidean space with $Q^{2}=\vec q^{\,2}+\omega_{n}^{2}$ and $Q_{0}=i\omega_{n}$:
\beq
M_n(Q)=
\begin{pmatrix}
 Q^2+m_s^2&2m\Delta^*&2m\Delta\\
 2m\Delta&\fra(Q^2+m_d^2)-2\mu Q_0&\Delta^2\\
 2m\Delta^*&\Delta^{*2}&\fra(Q^2+m_d^2)+2\mu Q_0
\end{pmatrix}.
\eeq
On the other hand, the matrix $N_n(Q)$ of the pion term is diagonal and proportional to the identity matrix as
\beq
N_n^{ij}=\delta^{ij}(Q^2+m_{\pi}^2) ~.
\eeq

The hadronic thermodynamical potential, $\Omega_{hadron}=-\frac{T}{V}\ln Z_{hadron}$, becomes
\beq
\label{ome}
\Omega_{hadron}&=&\fra\sum_{Q}\left(\ln\det M_n(Q)+\ln\det N_n(Q)\right)\nn
&=&\int\f{q}{3}\left[\sum_{i}^{3}\left(\frac{\omega_i}{2}+T\ln(1-e^{-\beta\omega_i})\right)
\right. \nn&&\left.
+ \,3\left(\frac{\omega_{\pi}}{2}+T\ln(1-e^{-\beta\omega_{\pi}})\right)\right]~.
\eeq
The pion excitation energy is $\omega_{\pi}(\vec q)=\sqrt{\vec{q}\,^2+m_{\pi}^2}$.  The excitation energies $\omega_i (i=1,2,3)$ are the eigenvalues of the matrix $M(\omega,\vec{q})$ in \ref{app2}, obtained by solving the dispersion relation $\det M(\omega, \vec{q})=0$ with respect to $\omega$ numerically.  The momentum integrals of these boson zero point energies are divergent.  We regularize the integrals by the cut-off momentum $\Lambda$ as in the case of the quarks and subtract the zero point energies in the vacuum, $T=\mu=0$.  We can verify that the zero point energy contribution vanishes when $|\Delta|=0$.  At finite $|\Delta|$ there appears a finite contribution to the thermodynamical potential.

A related study of the thermodynamical potential beyond the mean field approximation, including hadron contributions in the two-color NJL model by using the Gaussian approximation, has been performed by He  \cite{He2010}.  The effect of the pair mode in condensed matter has been investigated by Diener {\it et al.} \cite{Diener2008}.  This method is applied to the NJL model and the effect of the hadrons is estimated by a perturbation method for the quark density \cite{He2010}.  An expansion of the $N_{c}=3$ two-flavor PNJL model beyond mean field approximation has been reported in ref. \cite{Roessner2008}.

\section{Numerical results}

\subsection{Polyakov potential}

For the discussion of the equation of state of quark-hadron matter, it is important to take into account the quark confinement effect.  We adopt here the Fukushima method and add the Polyakov loop potential $U(\Phi[A],\Phi^\ast[A];T)$ to the two-color NJL Lagrangian~\cite{Fukushima2004}.  The derivative $\partial^{\mu}$ acting on the quark field is replaced by the covariant derivative $D^\mu=\partial^\mu-iA^\mu$.  The temporal background color gauge field is introduced $A_4=iA^{0}$ with $A^0=gA_a^0 \frac{t_a}{2}$ and the $SU(2)$ Pauli matrices $t_a (a=1,2,3)$ in color space.
In the Polyakov gauge this temporal gauge field is diagonal in the color space.  For the $SU(2)$ color group it is represented as $t_3\theta$ where $\theta$ is real.

The Polyakov loop potential $U$ is written as \cite{Brauner2009}
\beq
U(\Phi,T)=-bT(24 \Phi^2e^{-\beta a}+\ln(1-\Phi^2))~,
\eeq
in terms of the traced Polyakov loop
\beq
 \Phi=\frac{1}{N_c}\tr e^{i\beta A_4}=\cos(\beta\theta)~.
\eeq
The $\log$ term comes from the Jacobian in the $SU(2)$ color space.  The two parameters $a=$858.1MeV and $b^{1/3}=$210.5MeV are taken from the work of Brauner {\it et al.}~\cite{Brauner2009}.

Taking into account the Polyakov loop effect implies the replacement
\beq
\ln(1+e^{-\beta E_\Delta^+})\rightarrow \fra \ln(1+2\Phi e^{-\beta E_\Delta^+}+e^{-2\beta E_\Delta^+})~.
\eeq
Hence, the Fermi function $f(E)$ in the mean field equation is replaced by
\beq
 \tilde{f}(E)=\frac{1+\Phi e^{\beta E}}{1+2\Phi e^{\beta E}+e^{2\beta E}}~.
\eeq
This means that single quark quasiparticles are suppressed in the hadronic phase as $\Phi\rightarrow 0$, whereas color-singlet combinations with two quarks survive. 

We now proceed to discuss numerical results.  For this purpose we fix the parameters of the PNJL model following the paper of Brauner {\it et al.}, the list of which is given in Table I of~\cite{Brauner2009}.   These parameters are set such that the pion mass in free space is $m_\pi=140\,$MeV and the pion decay constant is $f_\pi=\sqrt{2/3}\times 93\,$MeV=76$\,$MeV, reflecting the $\sqrt{N_{c}}$ behavior of $f_{\pi}$.  The coupling constants are $G_{0}=H_{0}=7.23\,$GeV$^{-2}$, the cut-off momentum is $\Lambda= 657\,$MeV and the bare quark mass is $m_{0}=5.4\,$MeV.

\subsection{Phase diagram and order parameters}

\begin{figure}[htbp]
\hspace{3mm}
\begin{minipage}{0.45\hsize}
 \begin{center}
 \includegraphics[bb=0 0 360 252,scale=0.5,clip]{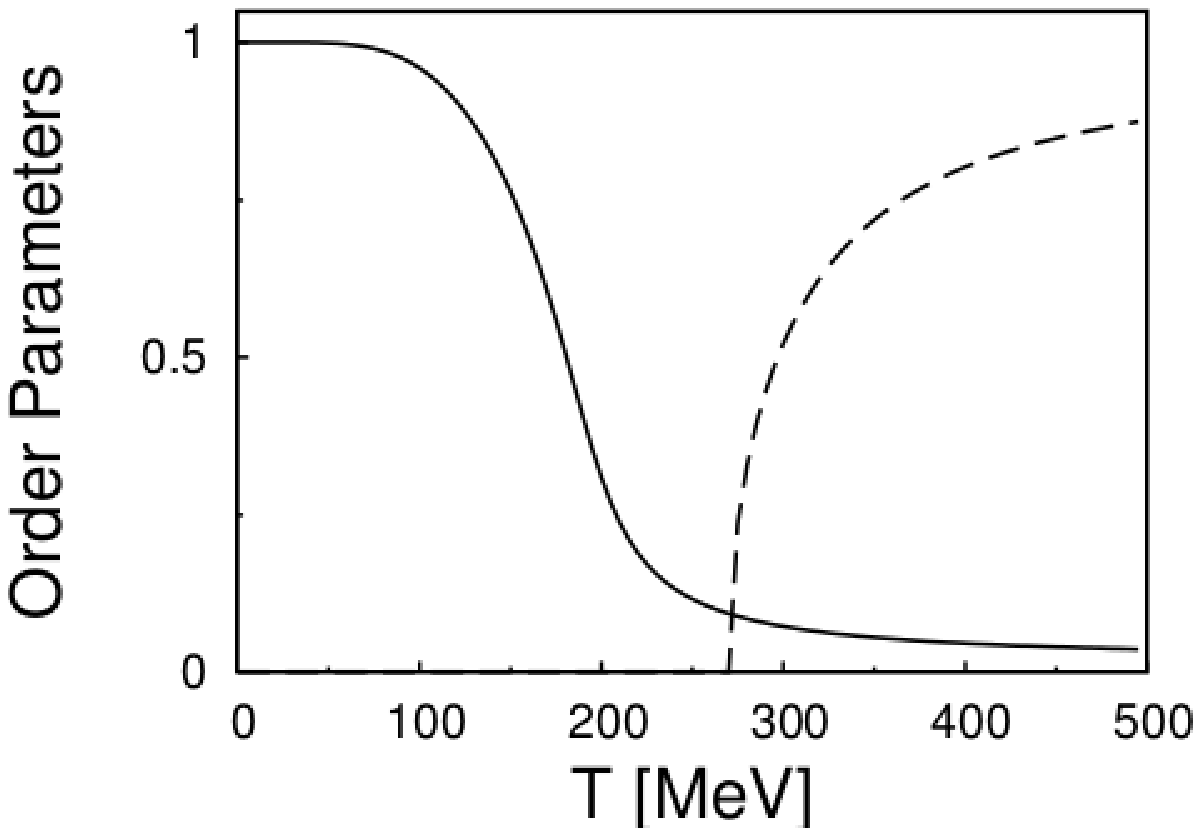}
\caption{\small The chiral condensate $\sigma_{0}$ in unit of $\sigma_0(T=0)$ at zero temperature (solid curve) and the Polyakov loop $\Phi$ (dashed curve) as functions of temperature at zero chemical potential ($\mu$=0) for the case of the NJL model with decoupled Polyakov loop.}
 \label{chiral_WOP}
 \end{center}
\end{minipage}
\hspace{3mm}
\begin{minipage}{0.45\hsize}
\begin{center}
\includegraphics[bb=0 0 360 252,scale=0.5,clip]{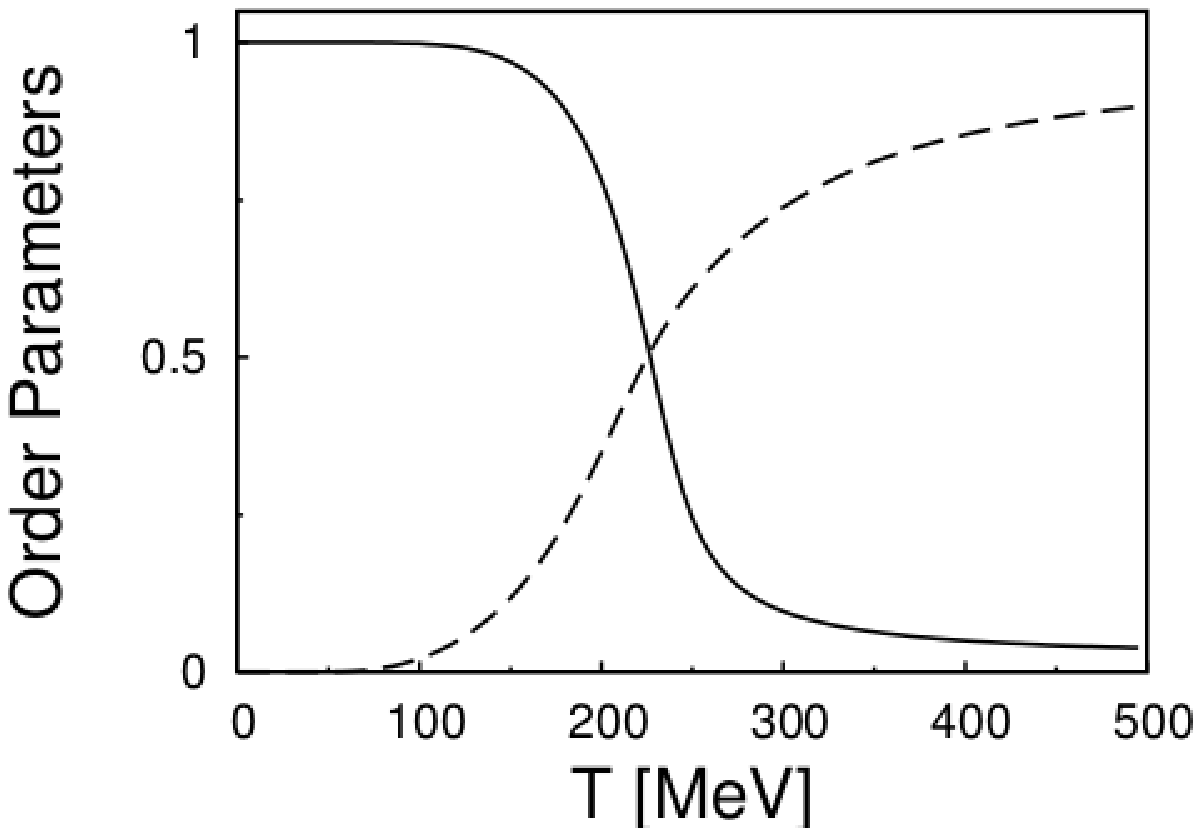}
\caption{\small The chiral condensate $\sigma_{0}$ (solid curve) and the Polyakov loop $\Phi$ (dashed curve) as functions of temperature at zero chemical potential ($\mu$=0) for the case of the PNJL model including coupling of quarks to the Polyakov loop.}
\label{chiral_WP}
\end{center}
\end{minipage}
\end{figure}

We start by showing the chiral condensate, the diquark condensate and the Polyakov loop with and without the Polyakov loop coupling effect.  These condensates are obtained by solving the gap equations for the chiral condensate $\sigma_{0}$ and the diquark condensate $|\Delta_{0}|$.  Results are presented in Figs. \ref{chiral_WOP} and \ref{chiral_WP} as functions of temperature at zero chemical potential ($\mu=0$).
The crossover temperature of the chiral condensate defined at $\sigma_c=\fra \sigma_0$ is $180\,$MeV.  The critical temperature of the Polyakov loop is set at $T_0=270\,$MeV for the pure gauge case without coupling to quarks.  At this point we adopt the $N_{c}=3$ value of $T_{0}$ as in ref. \cite{Brauner2009}.  According to ref. \cite{lucini04}, the $N_{c}$-dependence of $T_{0}$ goes approximately as $T_{0}/\sqrt{\sigma} \simeq 0.6+0.45/N_{c}^{2}$, suggesting roughly a 10\% difference between $T_{0}(N_{c}=2)$ and $T_{0}(N_{c}=3)$ that we can ignore.  When the coupling between the quark and the Polyakov loop is introduced in the PNJL model, the chiral and deconfinement crossover transition temperatures become strongly correlated as shown in Fig. \ref{chiral_WP}.  The crossover temperatures of both order parameters defined at half of their full values are now $T_\chi\simeq T_{dec}\simeq 225\,$MeV.

\begin{figure}[htbp]
\hspace{3mm}
\begin{minipage}{0.45\hsize}
 \begin{center}
 \includegraphics[bb=0 0 360 252, scale=0.5]{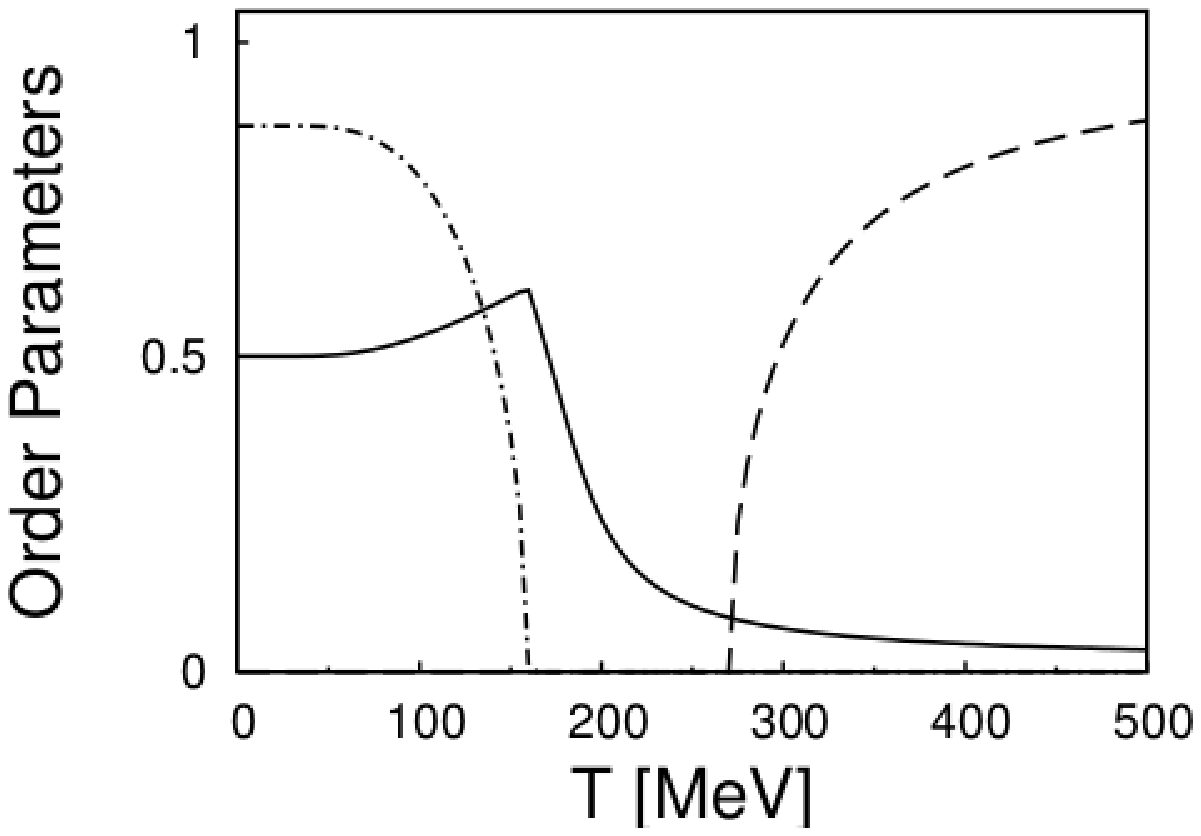}
 \caption{\small The chiral condensate $\sigma_{0}$ (solid curve), the diquark condensate $|\Delta_{0}|$ (dash-dotted curve) in unit of $\sigma_0(T=0)$ and the Polyakov loop $\Phi$ (dashed curve) as functions of temperature at finite chemical potential ($\mu=100$MeV) for the case of the NJL with decoupled Polyakov loop.}
 \label{chiral_muWOP}
 \end{center}
\end{minipage}
\hspace{3mm}
\begin{minipage}{0.45\hsize}
 \begin{center}
 \includegraphics[bb=0 0 360 252, scale=0.5]{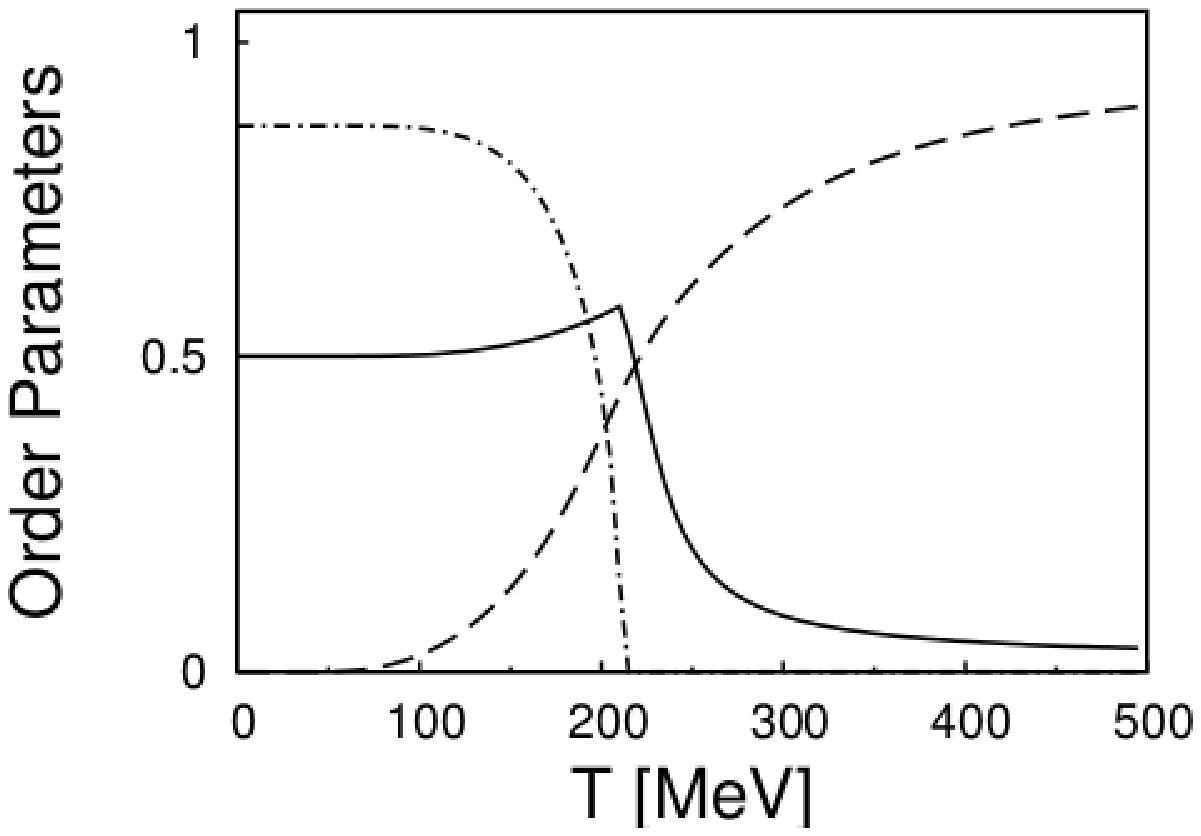}
 \caption{\small The chiral condensate $\sigma_{0}$ (solid curve), the diquark condensate $|\Delta_{0}|$ (dash-dotted curve) in unit of $\sigma_0(T=0)$ and the Polyakov loop $\Phi$ (dashed curve) as functions of temperature at finite chemical potential ($\mu=100$MeV) for the case of the PNJL model including coupling of quarks to the Polyakov loop.}
 \label{chiral_muWP}
 \end{center}
\end{minipage}
\end{figure}
Results at finite chemical potential ($\mu=100\,$MeV) are shown in Figs. \ref{chiral_muWOP} and \ref{chiral_muWP}.  In this case the diquarks condense and the chiral condensate is reduced smaller where the diquark condensate is finite.  Again the coupling of quarks to the Polyakov loop makes the deconfinement transition smooth, so that both chiral and deconfinement crossover temperatures become quite similar. 

\begin{figure}[htbp]
\hspace{3mm}
\begin{minipage}{0.45\hsize}
 \begin{center}
 \includegraphics[bb=0 0 360 252, scale=0.5]{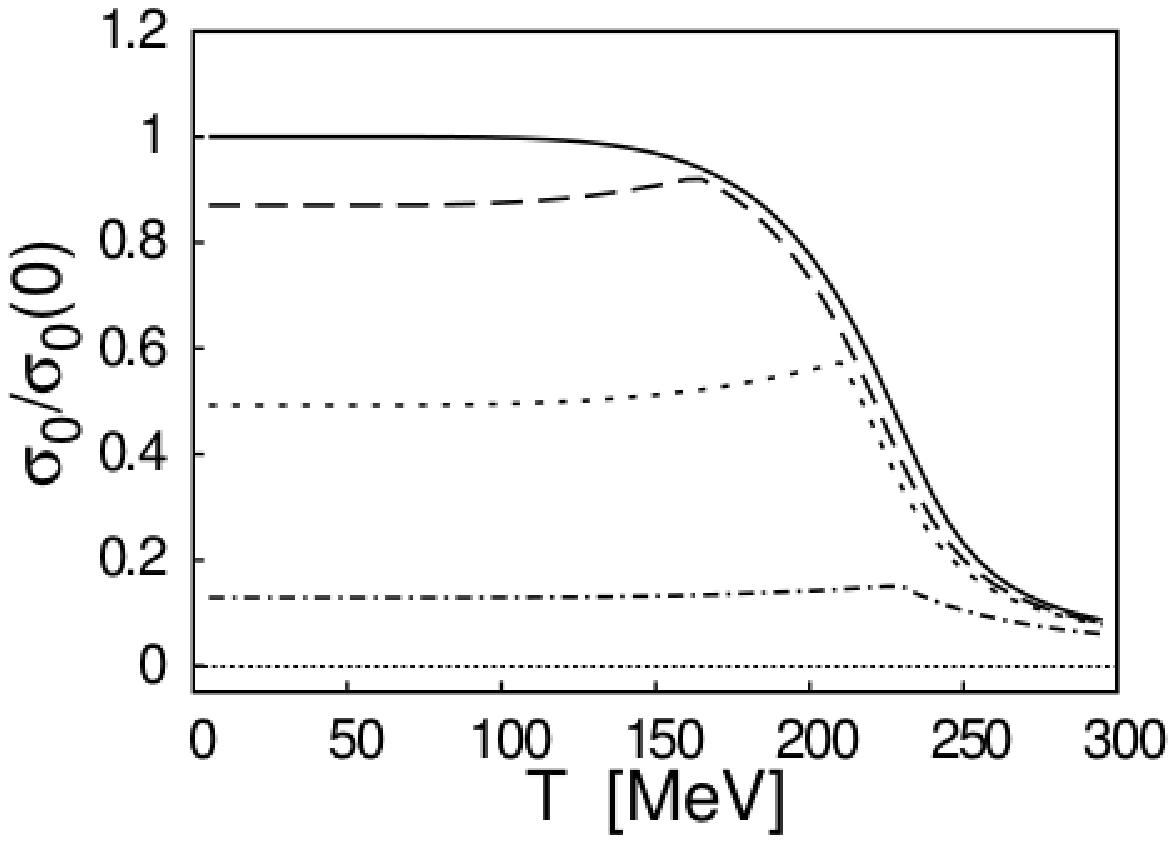}
 \caption{\small The chiral condensate $\sigma_{0}$ in units of $\sigma_0(T=0)$ for $\mu=0$, 75, 100 and 200$\,$MeV (from top) as functions of temperature.}
 \label{sigma}
 \end{center}
\end{minipage}
\hspace{3mm}
\begin{minipage}{0.45\hsize}
 \begin{center}
 \includegraphics[bb=0 0 360 252, scale=0.5]{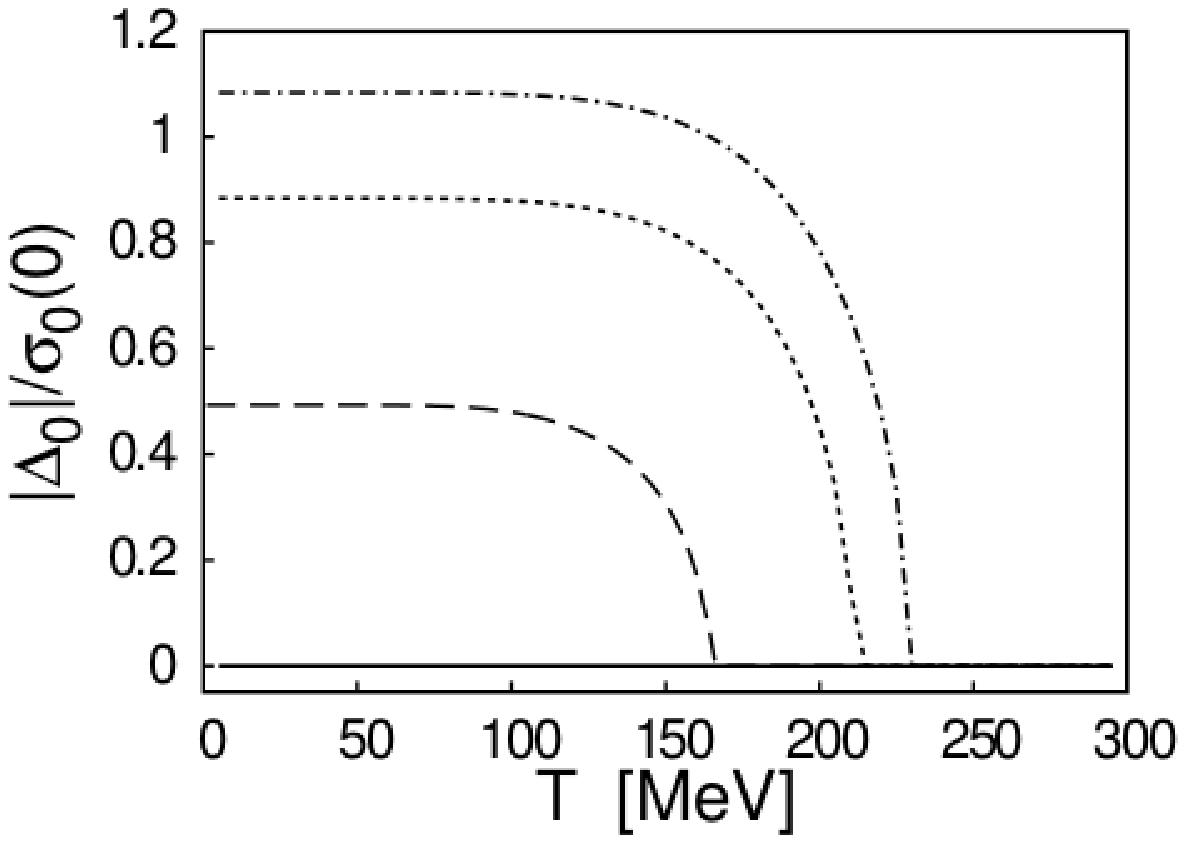}
 \caption{\small The diquark condensate $|\Delta_{0}|$ in units of $\sigma_0(T=0)$ for $\mu=0$, 75, 100 and 200$\,$MeV (from bottom) as functions of temperature.}
 \label{delta}
 \end{center}
\end{minipage}
\end{figure}

The phase boundary of the de-confinement transition ($\Phi \sim 0.5$) is insensitive to the chemical potential as discussed by Brauner {\it et al.}~\cite{Brauner2009}.  Hence, we plot in Fig. \ref{sigma} and \ref{delta} only $\sigma_{0}$ and $|\Delta_{0}|$ for various chemical potentials as functions of temperature.  The chiral condensate $\sigma_{0}$ shown by the top smooth curve ($\mu=0$) in Fig. \ref{sigma} stay unchanged until diquark condensate sets in at $\mu=\fra m_{\pi}=70\,$MeV.  $\sigma_{0}$ at $\mu=75\,$MeV is depleted in the small temperature region, where the diquark condensate $|\Delta_{0}|$ is finite as shown in the right hand figure.  This behavior continues as the chemical potential increases as shown for $\sigma_{0}$ and $|\Delta_{0}|$ at $\mu=100$ and $200\,$MeV~\cite{Kogut2000, Ratti2004}.  These behaviors agree with the results shown in Fig. 3 of ref. \cite{Brauner2009}.


\subsection{BCS-BEC crossover}

\begin{figure}[htbp]
\hspace{3mm}
\begin{minipage}{0.45\hsize}
 \begin{center}
 \includegraphics[bb=0 0 360 252, scale=0.5]{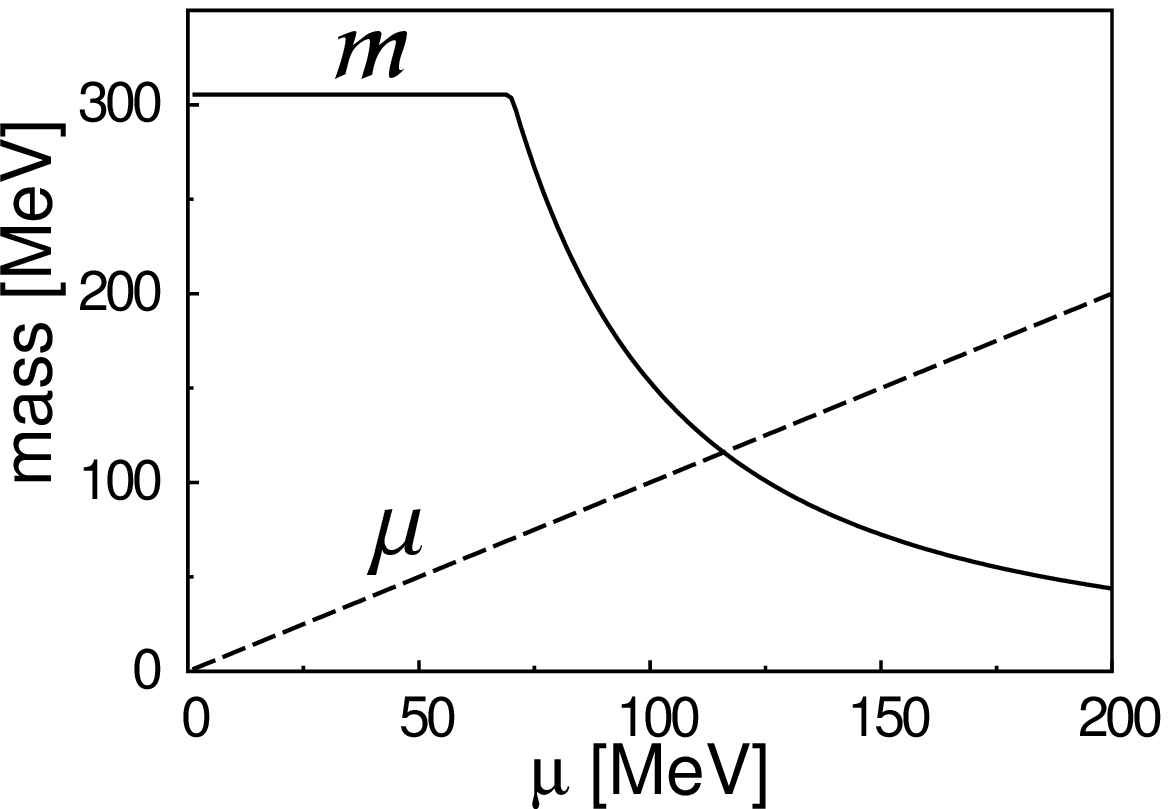}
 \caption{\small The dynamical quark mass $m$ (solid curve) and the chemical potential $\mu$ (dashed curve) as functions of the chemical potential.  The BEC-BCS crossover point corresponds to the crossing point, $\mu =120\,$MeV.}
 \label{becbcs}
 \end{center}
\end{minipage}
\hspace{3mm}
\begin{minipage}{0.45\hsize}
 \begin{center}
 \includegraphics[bb=0 0 360 252, scale=0.5]{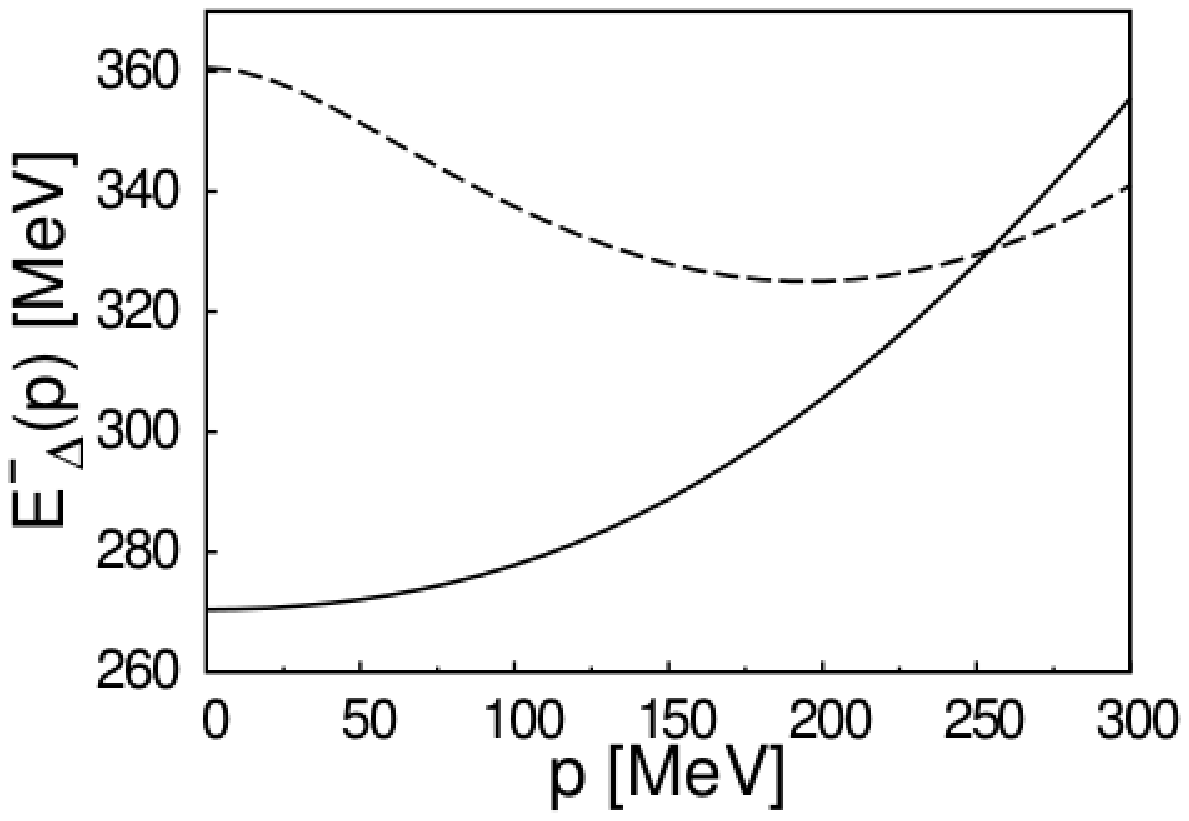}
 \caption{\small The quasi-particle energy of quarks at $\mu=100\,$MeV (solid curve) and at $\mu=200\,$MeV (dashed curve) as functions of momentum $p$.}
 \label{qpe}
 \end{center}
\end{minipage}
\end{figure}

The two gap equations (\ref{gap}) and (\ref{gapdelta}) for $\sigma_{0}$ and $|\Delta_{0}|$ point to an interesting interplay concerning the BEC-BCS crossover.  When the chemical potential $\mu$ exceeds the critical point of diquark condensation, the dynamical quark mass $m=g_{0}\sigma_{0}+m_{0}$ is much larger than the chemical potential $\mu$ as shown in Fig. \ref{becbcs}.  Hence, the system is in the BEC phase.  This is demonstrated in Fig. \ref{qpe} showing the quasi-particle energy of quarks, $E_{\Delta}^{-}(\vec p\,)=((E_{p}^{-})^{2}+g_{d}^{2}|\Delta_{0}|^{2})^{1/2}$, with $E_{p}^{-}=\sqrt{m^{2}+\vec p^{\,2}}-\mu$ at $\mu=100\,$MeV.  At this chemical potential the minimum of the quasi-particle energy is at zero momentum.   As $\mu$ increases, the gap $|\Delta_{0}|$ increases, while the dynamical quark mass $m$ decreases as shown in Fig. \ref{becbcs}.  Once the chemical potential $\mu$ exceeds the dynamical quark mass, $\mu \geq m$, the system undergoes the BEC-BCS crossover and turns into the BCS phase as $\mu$ increases further.  The quasi-particle energy $E_{\Delta}^{-}$ at $\mu=200\,$MeV shown in Fig. \ref{qpe} has a minimum at finite momentum.  The crossover point is insensitive to the temperature (see Figs. \ref{sigma} and \ref{delta}) as long as the diquark condensation is finite.

\subsection{Meson and diquark masses}

\begin{figure}[htbp]
\hspace{3mm}
\begin{minipage}{0.45\hsize}
 \begin{center}
 \includegraphics[bb=0 0 360 252, scale=0.5]{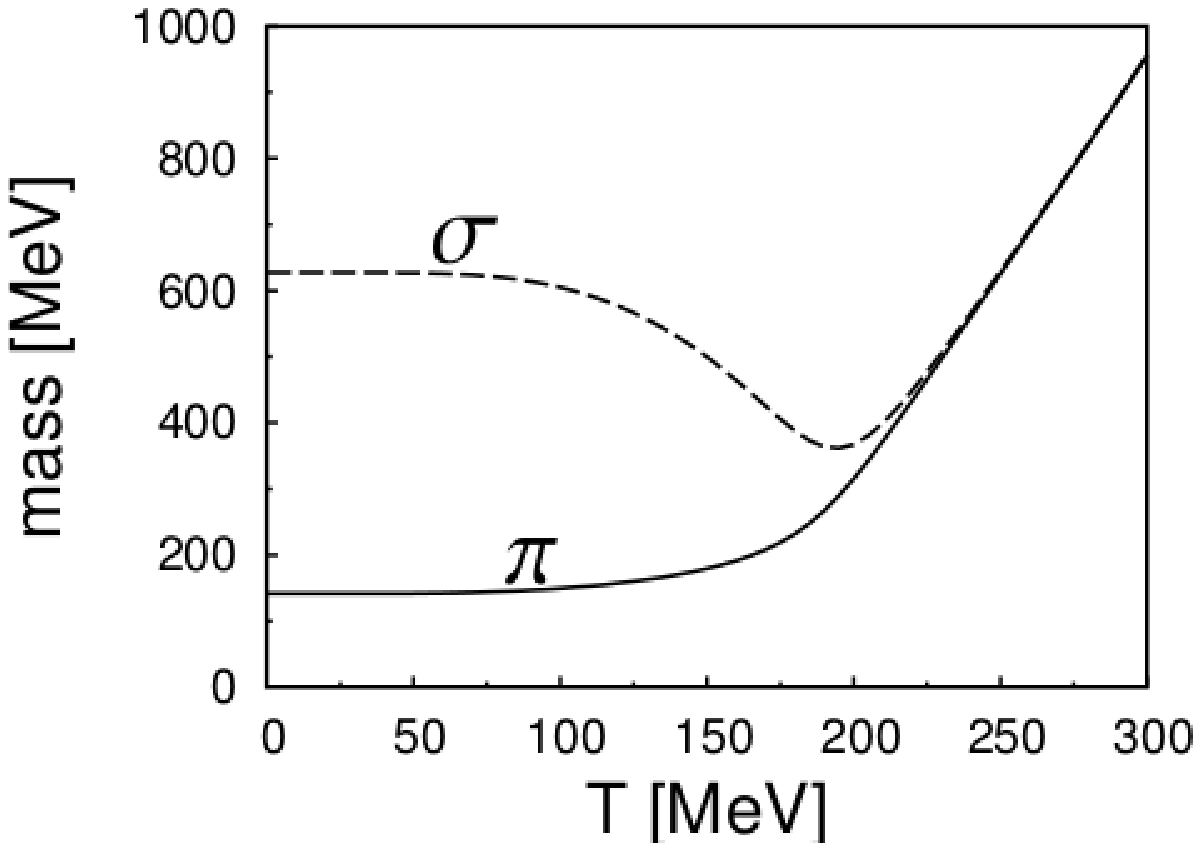}
 \caption{\small Sigma (dashed curve) and pion masses (solid curve) in MeV as functions of temperature in MeV at zero chemical potential ($\mu$=0) for the case of the NJL model with decoupled Polyakov loop.}
 \label{mesonmass_WOP}
 \end{center}
\end{minipage}
\hspace{3mm}
\begin{minipage}{0.45\hsize}
 \begin{center}
 \includegraphics[bb=0 0 360 252, scale=0.5]{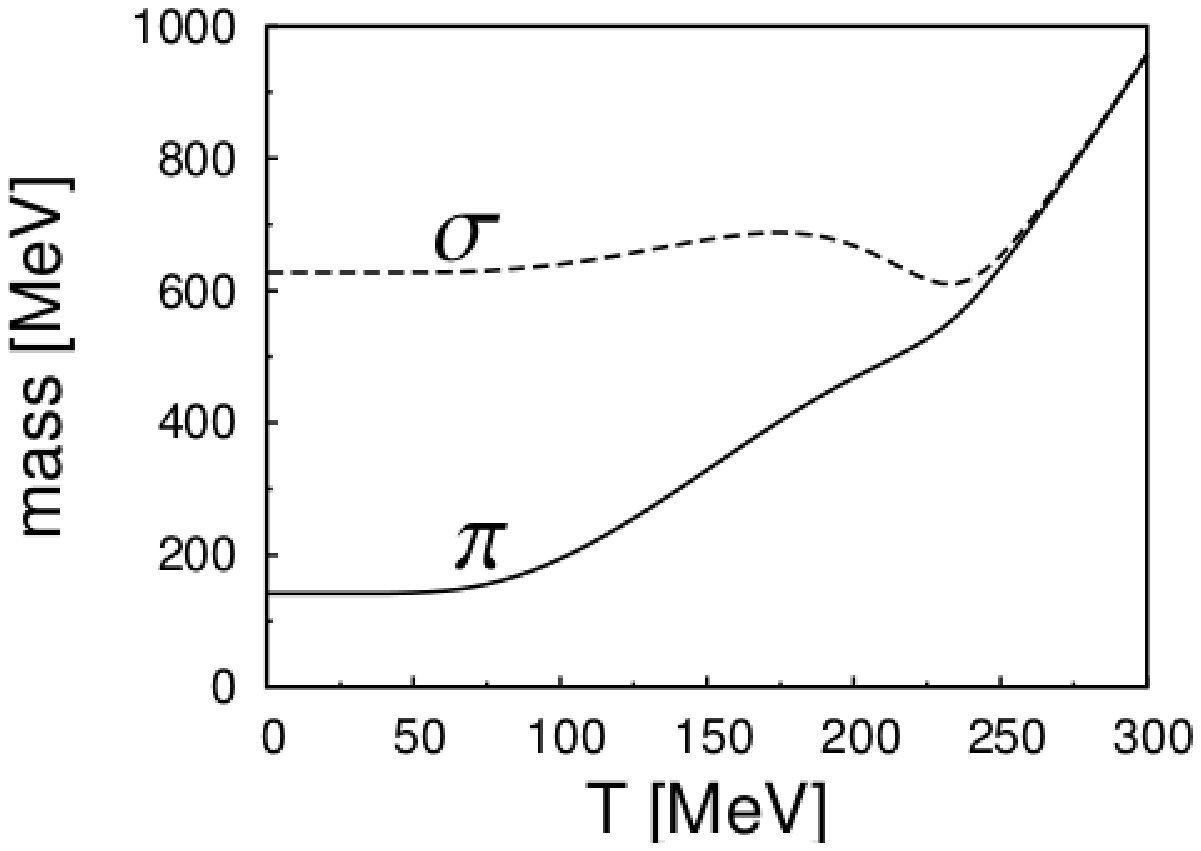}
 \caption{\small Sigma (dashed curve) and pion masses (solid curve) in MeV as functions of temperature at zero chemical potential ($\mu$=0) for the case of the PNJL model with coupling of quarks to the Polyakov loop.}
 \label{mesonmass_WP}
 \end{center}
\end{minipage}
\end{figure}

We start with scalar sigma boson and pion masses at zero chemical potential.  In this case, the diquark condensate is zero and there is no sigma-diquark mixing.  The behavior of these masses is similar to that found in the $SU(3)$ color NJL model.  The only difference is the role of the Polyakov loop for two colors here as compared to the standard color $SU(3)$ case.  We show the sigma and pion masses as functions of temperature with and without the quark-Polyakov-loop coupling at zero chemical potential ($\mu$=0) in Figs. \ref{mesonmass_WOP} and \ref{mesonmass_WP}.  The explicit chiral symmetry breaking by the small quark mass, $m_0=5.4\,$MeV, gives the Nambu-Goldstone pion a small mass, $m_{\pi}=140\,$MeV at zero temperature, while the sigma mass, $m_{s}=610\,$MeV stays around twice the dynamical quark mass.  For zero chemical potential, the diquark-baryon masses are degenerate with the pion mass due to the PG symmetry.  As the temperature increases, the pion mass starts to increase, while the sigma mass decreases until the temperature approaches the crossover temperature of about 180$\,$MeV.   The two masses meet at the crossover temperature and increase jointly as shown in Fig. \ref{mesonmass_WOP}.

When the quark-Polyakov-loop coupling is introduced, the pion mass increases earlier with temperature as shown in Fig. \ref{mesonmass_WP}.  The reason is that the quark mass drops more slowly with increasing temperature as shown in Fig. \ref{chiral_WP}.  The sigma mass stays almost constant up to the  crossover temperature and then both the pion and sigma masses increase rapidly with increasing temperature.

\begin{figure}[htbp]
\hspace{3mm}
\begin{minipage}{0.45\hsize}
 \begin{center}
 \includegraphics[bb=0 0 360 252, scale=0.5]{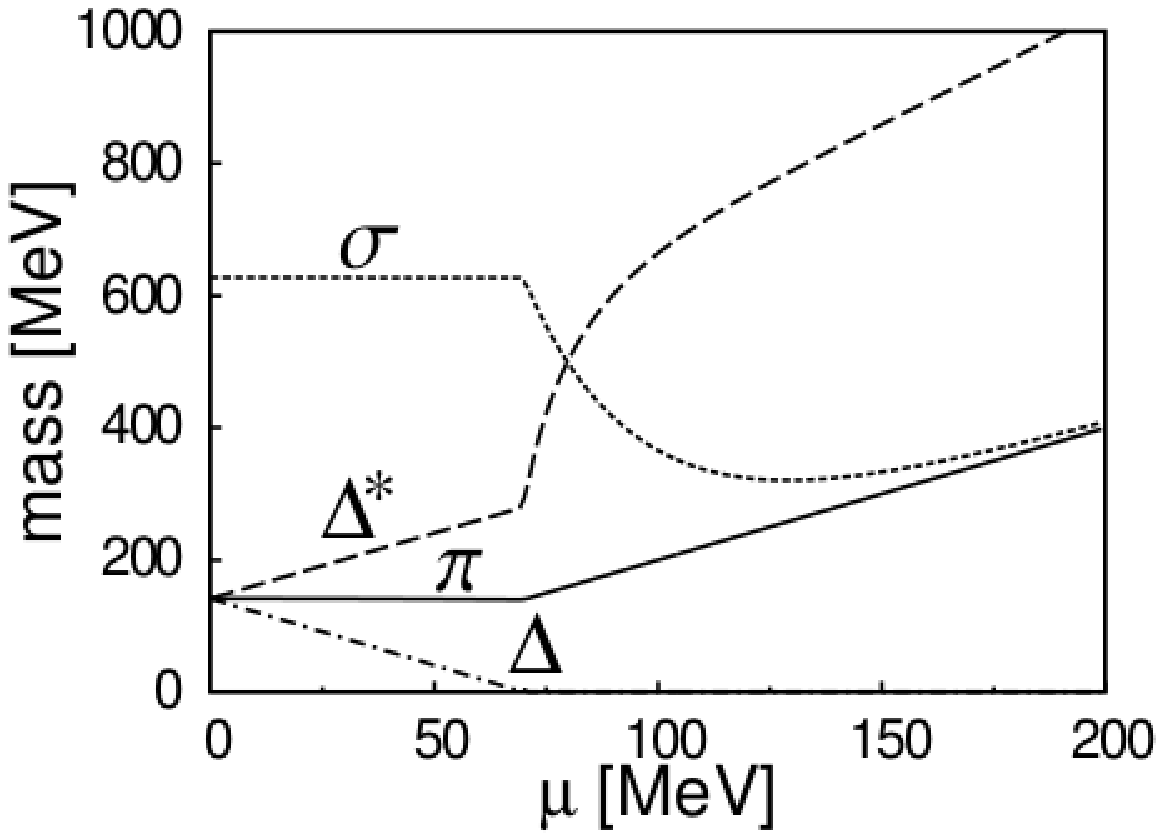}
 \caption{\small The sigma (dotted curve), the pion masses (solid curve) and diquark masses (dash-dotted and dashed curves) in MeV as functions of the chemical potential $\mu$ at zero temperature $(T=0)$.  The sigma and diquark coupling is dropped when $|\Delta_{0}| \ne 0$.}
 \label{hadronmass_wom}
 \end{center}
\end{minipage}
\hspace{3mm}
\begin{minipage}{0.45\hsize}
 \begin{center}
 \includegraphics[bb=0 0 360 252, scale=0.5]{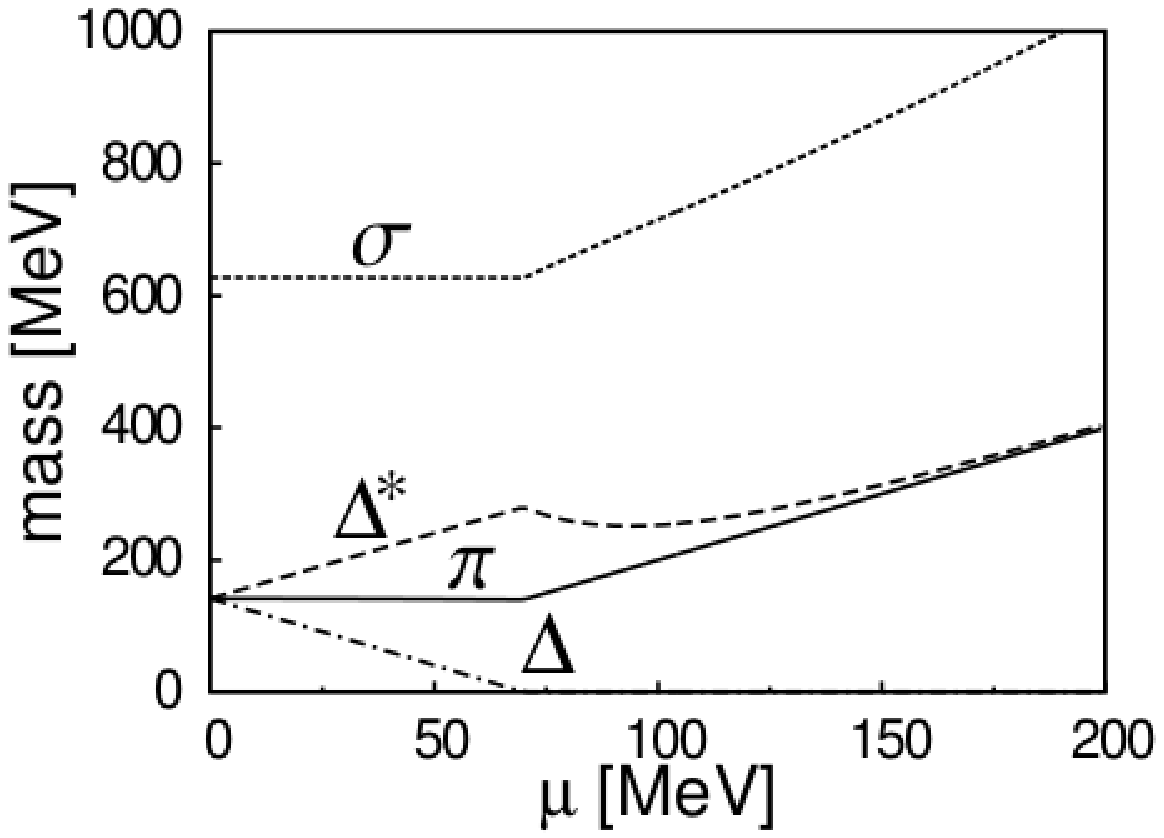}
 \caption{\small The sigma (dotted curve), the pion masses (solid curve) and diquark masses (dash-dotted curve) and (dashed curve) in MeV as functions of the chemical potential $\mu$ at zero temperature $(T=0)$.  The sigma and diquark coupling is introduced when $|\Delta_{0}| \ne 0$.}
 \label{hadronmass}
 \end{center}
\end{minipage}
\end{figure}
In the case of diquark condensation at finite chemical potential $\mu > \frac{m_{\pi}}{2}$, all scalar hadronic degrees of freedom have non-diagonal coupling terms.  The mass matrix therefore needs to be diagonalized together with the chemical potential terms.  We discuss the behavior of the masses after diagonalization in Appendix B.  The renormalized pion mass is written as
\beq
m_{\pi}^2&=(M_{s}^2-2g_0^2I_2+4g_0^2I_0\mu^2)/(g_{0}^{2}I_0)=\frac{\varepsilon}{\sigma_{0}}~.
\eeq
The pion mass stays constant at 140$\,$MeV and proportional to the square root of the bare quark mass until the onset of diquark condensation.  The sigma mass is larger than the pion mass, $m_{s}^{2}=m_{\pi}^{2}+4m^{2}$, with $m$ the dynamical quark mass as discussed in Appendix B.  
Meson and the diquark-baryon masses are shown as functions of the chemical potential $\mu$ at zero temperature in Figs. \ref{hadronmass_wom} and \ref{hadronmass}.  The diquark masses equal the pion mass at $\mu$=0 due to the Pauli-G\"{u}rsey symmetry.  At $\mu>0$  one of the diquark mass decreases linearly and the other one increases linearly with $\mu$ as discussed in Appendix B.  This is due to the trivial fact that there exist diquark and antidiquark states with energies at rest $E_{d}=\pm M_d=\pm140\,$MeV.  Their excitation energies change linearly with $\mu_{B}=2\mu$, but with opposite signs. 

After diquark condensation, which occurs at $\mu=\fra m_\pi$ ($\mu_{B}=m_{\pi}$), the pion mass increases linearly with $\mu$ as discussed in Appendix B.   As for the sigma and diquark masses, they couple after diquark condensation, $|\Delta_{0}|\neq 0$.  The mass matrix is
\beq
\label{sdmat}
 M(\omega,\vec{q})=
\begin{pmatrix}
  {q}^2-m_s^2&-2\Delta^* m& -2\Delta m\\
 -2\Delta m&\fra({q}^2-m_d^2)+2\mu\omega&-\Delta^2\\
 -2\Delta^* m&-\Delta^{*2}&\fra({q}^2-m_d^2)-2\mu\omega
\end{pmatrix}~.
\eeq
where we have written $q^{2}=\omega^{2}-\vec q^{\,2}$.  The mass becomes
\beq
m_{d}^{2}=(M_{d}^2-2g_d^2I_2+2g_d^2I_0|\Delta|^2)/(g_{0}^{2}I_0)
 =2 |\Delta|^2 ~.
\eeq
The mass spectra for sigma, diquark and antidiquark are obtained by solving the dispersion relation $\det M(\omega,0)=0$ with respect to $\omega$.  One solution is $\omega^{2}=0$, and we are left with a 2nd order equation for $\omega^{2}$, which is solved analytically as
\beq
\omega^2=2 |\Delta|^2+10\mu^2+2m^2\pm\sqrt{(6\mu^2+2 |\Delta|^2+2m^2)^2-48\mu^2m^2}~.
\eeq
We show in Fig. \ref{hadronmass} the results of the diagonalized masses.  The coupling of the sigma meson and the diquark-baryons is very large.  When the coupling is suppressed the sigma mass drops slightly as $\mu$ approaches the crossover chemical potential $\mu \sim 150\,$MeV and then increases together with the pion mass as shown in Fig. \ref{hadronmass_wom}.  At the same time, the antidiquark mass increases rapidly with the chemical potential after diquark condensation.  On the other hand, in the case of strong scalar-diquark coupling (Fig. \ref{hadronmass}), the quantum-mechanical ``non-crossing'' rule is at work.  The sigma mass increases continuously and the antidiquark mass joins the slowly increasing pion mass.

\subsection{Equation of state of quark-hadron matter}

Next, consider the equation of state (EOS) of quark-hadron matter at various chemical potentials $\mu$ as functions of the temperature $T$.   The pressure is given by the thermodynamical potential, $p=-\Omega$.  For the case of only quarks, $\Omega=\Omega_{MF}$ (\ref{omegamean1}), the pressure is essentially zero as shown by the thin solid curve in the confined region (small temperature), and gradually increases with temperature as deconfinement sets in featuring the entanglement of chiral symmetry and Polyakov loop effects.  Finally the pressure becomes large at high temperature above the crossover transition.  
\begin{figure}[htbp]
\hspace{3mm}
\begin{minipage}{0.45\hsize}
 \begin{center}
 \includegraphics[bb=0 0 360 252, scale=0.5]{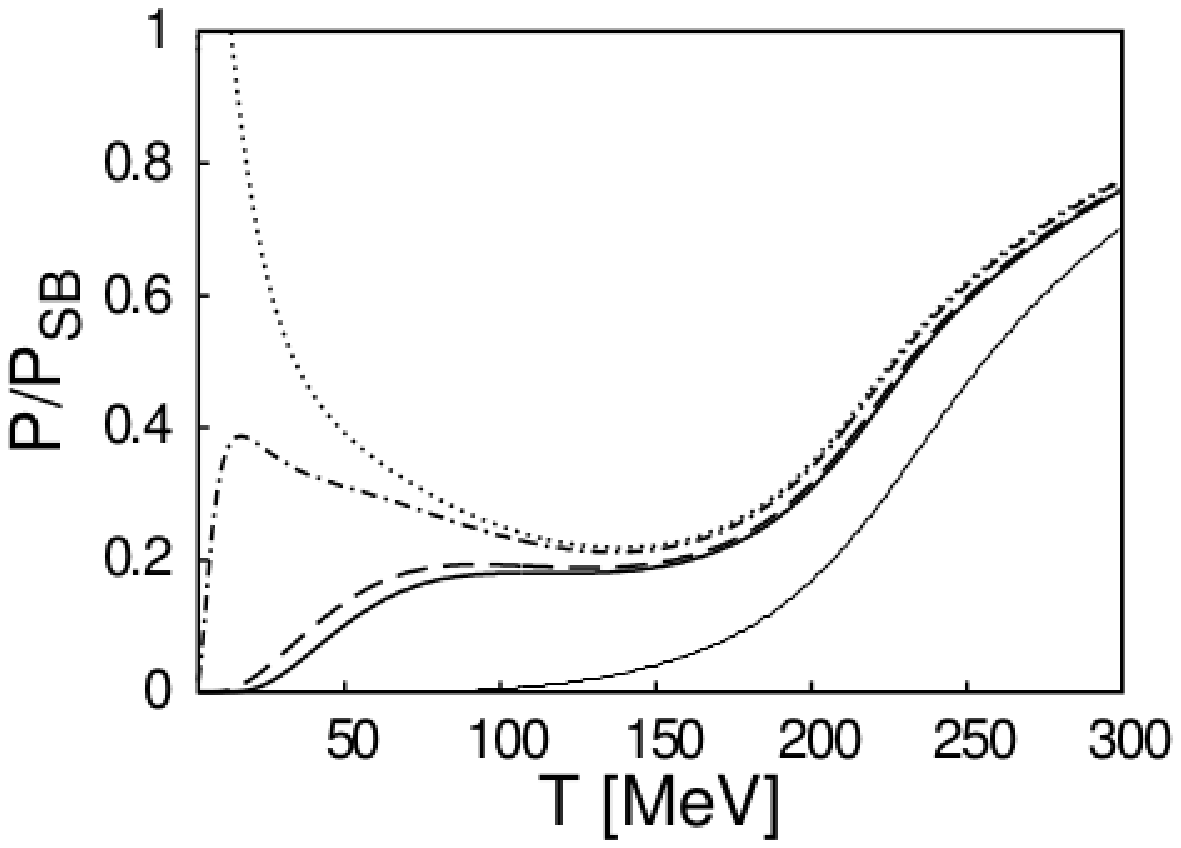}
 \caption{\small The pressure divided by the Stefan-Boltzmann pressure for various chemical potentials $\mu=$0 (solid), 30 (dashed), 60 (dash-dotted) and 66$\,$MeV (dotted) below diquark condensation as functions of temperature.  Shown also is the one with only the quark degree of freedom (thin solid curve).}
  \label{eos_wop}
 \end{center}
\end{minipage}
\hspace{3mm}
\begin{minipage}{0.45\hsize}
 \begin{center}
 \includegraphics[bb=0 0 360 230, scale=0.53]{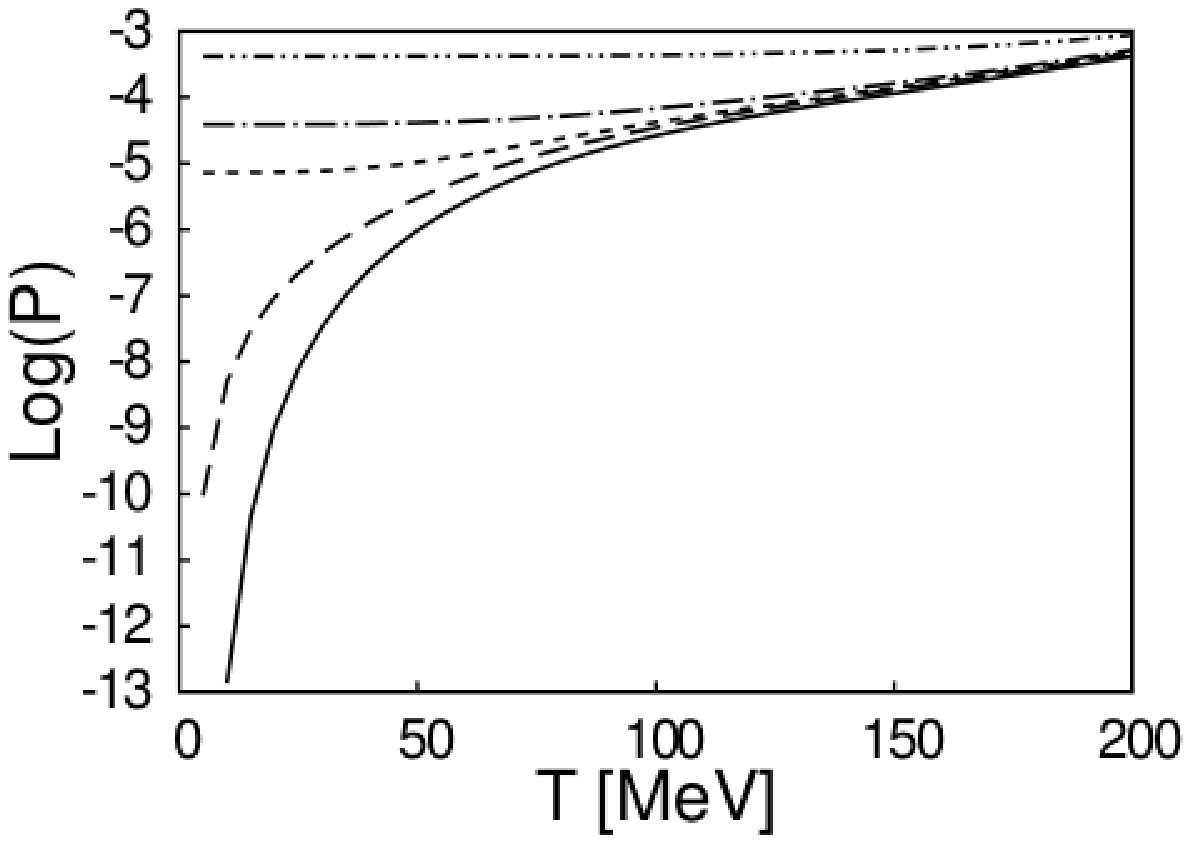}
 \caption{\small The logarithm of the pressure $\log(P)$ for various chemical potentials for $\mu=0$, $60\,$MeV below $\mu_{c}=70\,$MeV and for $\mu=80$, 100 and $200\,$MeV above $\mu_{c}$ as functions of temperature.  The unit of pressure is GeV$^{4}$.}
  \label{eoslog}
 \end{center}
\end{minipage}
\end{figure} 

In the confined region, the essential active degrees of freedom are hadrons.  Their dynamics is governed by the hadron Lagrangian derived previously, in which mesons and diquark-baryons interact.  We calculate thermodynamical potential, taking only the mass terms and integrating out the hadron fields by the Gaussian approximation as explained in Sect.$\,$3.   As mentioned previously, the divergent zero-point energy terms in Eq. (\ref{ome}) are regularized by introducing the cut-off momentum $\Lambda$.  The zero point energy at $T=0$ and $\mu=0$ is then subtracted from the thermodynamical potential.  For the case $\Delta=0$, this procedure makes the contribution of the zero-point energy terms vanish.  Hence, the whole contribution to the pressure from the hadrons comes from the temperature dependent terms for $\Delta=0$.  

Shown in Fig. \ref{eos_wop} is the pressure at $\mu=0$, 30, 60 and 66$\,$MeV.  Given the small pion and diquark-baryon masses, the contributions of these degrees of freedom make the pressure significantly different from that with quark degrees of freedom only.  This result is qualitatively similar to the one of the color $SU(3)$ PNJL model \cite{Roessner2008}, but here, due to the additional pressure of the diquark-baryon fields, the effect of the hadron contributions is much larger.  The pressure increases rapidly as the chemical potential approaches the critical chemical potential $\mu_{c}=m_{\pi}/2$.  This rapid increase of the pressure is caused by the diquark-baryon mode whose energy drops as $\omega_{d}=\omega_{\pi}-2\mu$ with increasing $\mu$.  When the temperature becomes small ($T\sim 10\,$MeV), we can expand the logarithm and the pressure can be written approximately as 
\beq
p\simeq T \int \frac{d^{3}q}{(2\pi)^{3}}e^{-(\omega_{\pi}(\vec q)-2\mu)/T}\sim T \int_{0}^{q_{max}} \frac{q^{2}dq}{2\pi^{2}}e^{-(\omega_{\pi}(\vec q)-2\mu)/T}~,
\eeq
where $q_{max}$ is appreciable only for the diquark-baryon mode (would be zero mode) as $\mu \rightarrow \mu_{c}$.
Since the pressure is divided by the Stefan-Boltzmann pressure, the pressure ratio close to the critical temperature shows rapid growth when $\mu$ approaches the critical chemical potential.

Consider now the pressure over a wide range of chemical potentials $\mu$, in particular above the critical $\mu_{c}$ for diquark condensation.  The pressure is then dominated by quarks through their zero point motion renormalized by the vacuum value, with additional contributions from the zero point motion of mesons and diquark-baryons.  Given that these zero point motion effects do not vanish at zero temperature, it is more appropriate to present the pressure as such, not divided by the Stefan-Boltzmann pressure, on a logarithmic scale.  Results are shown in Fig. \ref{eoslog}.   The zero point motion is largely influenced by the presence of the Bogoliubov spectrum of the diquark zero mode, discussed in Appendix B~\cite{Brauner2008, He2010, Diener2008, Engelbrecht1997}.   The pressure curves in Fig. \ref{eoslog} are displayed in the whole $\mu$ range as functions of temperature.  Shown in Fig. \ref{eoslog} are the pressures at $\mu=$0, 60$\,$MeV below $\mu_{c}=m_{\pi}/2$ and $\mu=80$, 100 and $200\,$MeV above $\mu_{c}$.  The pressure below the critical chemical potential  drops to zero at $T=0$, while the pressure at $\mu > \mu_{c}$ stays finite at zero temperature.  The pressure at low temperature increases rapidly across the critical chemical potential.  The pressure at high temperature is dominated by the de-confined quark contribution and insensitive to the chemical potential.

\subsection{Baryon density}

An interesting point of comparison with color $SU(2)$ lattice simulations concerns the quark density as a function of chemical potential $\mu$ at zero temperature.  When the baryon number symmetry is broken, the diquark condensate becomes finite and the diquark-baryon becomes a Nambu-Goldstone boson.  From the onset of diquark condensation, the quark number becomes finite.  The quark density derived from the thermodynamical potential at mean field level is:
\beq
 \rho_{MF}=-\frac{\partial\Omega_{MF}}{\partial\mu}= \tr \int \f{p}{3} \left[\frac{E_p^+}{E_\Delta^+}-\frac{E_p^-}{E_\Delta^-}\right].
\eeq
Here, we write only the zero point oscillation terms, dropping the temperature dependent terms.
\begin{figure}[htbp]
\hspace{3mm}
 \begin{center}
 \includegraphics[bb=0 0 360 252, scale=0.6]{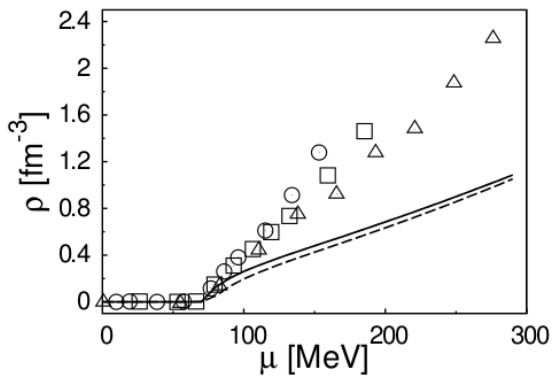}
 \caption{\small The quark density in unit of fm$^{-3}$ as a function of the chemical potential $\mu$ in MeV.  Shown by various points are the results of $SU(2)$ lattice simulations~\cite{Hands2001}.  The dashed curve denotes the result of the mean field approximation and the solid curve the results of the mean field and Gaussian approximation of the hadron contributions.}
 \label{density}
 \end{center}
\end{figure}

In Fig. \ref{density} the quark density is presented as a function of the chemical potential $\mu$.  As compared with lattice simulations~\cite{Hands2001} the mean field quark density comes out to be smaller than the lattice results by about a factor of two.  The quark density becomes non-zero when the diquark condensate develops, starting at $\mu=\fra m_\pi$.    The quark density increases with the chemical potential. 

Going beyond mean field level implies adding hadronic contributions to the quark density.  In a first step we use the Gaussian approximation, dropping higher order terms of the hadron Lagrangian, and integrate out the meson and diquark-baryon fields.  The hadronic part of the density is found by taking the derivative of $\Omega_{hadron}$, Eq. \eqref{ome}, with respect to the chemical potential $\mu$:
\beq
\rho_{hadron}=-\frac{\partial \Omega_{hadron}}{\partial \mu}=
-\frac{\partial }{\partial \mu} \int \f{q}{3} \left[\sum_{i}^{3}\frac{\omega_i(\mu)}{2}+\frac{3}{2}\omega_{\pi}(\mu)\right]~,
\eeq
where $\omega_{i}~(i=\sigma,\,d,\,d^{*})$ and $\omega_{\pi}$ are again the scalar boson, diquark, antidiquark and pion energies, respectively.  Only the temperature independent terms are written.  Note that $\rho_{hadron}$ vanishes for $\Delta=0$, while it becomes finite in the diquark condensed phase.  The zero point energies are calculated numerically and the momentum integrals are performed introducing the NJL cut-off $\Lambda=657\,$MeV.  We varied the cut-off momentum to 2$\,$GeV, and checked that the density increases only slightly.
The result is shown in Fig. \ref{density} by the solid curve.  Evidently, using the Gaussian approximation, the effect of the hadron fields on the quark density is very small.  This was anticipated by He \cite{He2010} in their analysis of the hadron contributions.

As already mentioned, the Gaussian approximation misses important hadronic interaction terms generated by the higher order pieces, $\mcl{L}^{(3)}$ and $\mcl{L}^{(4)}$, of the hadron effective Lagrangian.  These interactions include, for example, scalar boson exchange between diquarks and mesons in various possible combinations.  The strength of these couplings is controled by constant $\lambda$ in Eq. (\ref{lambda}).  We find $\lambda=33$ with the present parameter set.  Altogether, the net attraction provided by such mechanisms is expected to increase the density significantly as a function of $\mu$.  A systematic treatment of these effects is relegated to separate forthcoming work, using the Gaussian functional method discussed by Nakamura and Dmitrasinovic~\cite{Nakamura2001}.

\section{Summary and Conclusions}

In the present work we have investigated two-color quark-hadron matter at finite temperature and baryon chemical potential with the primary aim of studying, in a systematic and consistent way, the emergence and dynamics of baryonic degrees of freedom in addition to mesonic modes.  Baryons are realized as diquarks in $N_{c}=2$ QCD.  Their role as bosons is, of course, fundamentally different from the $N_{c}=3$ case in which baryons are fermions subject to the Pauli exclusion principle.  Nonetheless, the $N_{c}=2$ case is of conceptual interest because corresponding lattice simulations are not limited by the sign problem at non-zero, real chemical potential $\mu$. 

The modeling is performed using the NJL approach supplemented by Polyakov loop dynamics.  A full bosonization leads to an extended linear sigma model incorporating diquark degrees of freedom in the effective Lagrangian.  The thermodynamical potential derived from this effective Lagrangian is then studied in the mean field approximation and beyond.  Hadron properties are worked out as functions of temperature and chemical potential, and the phase diagram of quark-hadron matter is examined.  The behavior of chiral and diquark condensates and their intertwining at finite $\mu$, reflecting the underlying Pauli-G\"ursey symmetry, is discussed. 

The Polyakov loop plays an important role at finite temperature. The deconfinement transition is now correlated with the chiral condensate.  The characteristic temperatures for the chiral ($T_{\chi}$) and deconfinement ($T_{dec}$) cross-overs become about equal ( $T_{\chi}\simeq T_{dec} \simeq 225\,$MeV) once the quarks are coupled to the Polyakov loop.  At finite chemical potential and low temperature, $T<T_{dec}$, the diquark condensate plays an important role.  A non-zero diquark condensate emerges at $\mu=\fra m_\pi$, the onset of Bose-Einstein condensation in this model.  As $\mu$ increases the diquark condensate grows quickly and stays finite at large chemical potential.  With increasing diquark condensate, the chiral condensate decreases subject to the condition $\sigma_{0}^{2}+|\Delta_{0}|^{2}\simeq const$.

The meson and diquark masses behave naturally in accordance with the symmetry breaking pattern associated with chiral and diquark condensation.  The pion mass first stays constant and then grows linearly with the chemical potential once diquark condensation starts.  The sigma boson mass is about twice the dynamical quark mass until the onset of diquark condensation.  Due to the strong coupling of the sigma boson with diquark-baryons, the mixed sigma-diquark modes behave quite differently from the case without coupling. 

The equation-of-state (EOS) of the quark-hadron system has interesting properties as a function of increasing chemical potential.  When $\mu$ approaches its critical value $\mu_{c}=1/2\, m_{\pi}$ from below, the pressure divided by its Stefan-Boltzmann limit increase rapidly at low temperature.  This reflects the fact that the energy of the lowest diquark mode drops as $m_{\pi}-2\mu$ and hence contributes prominently to the pressure.  It becomes a zero mode at $\mu=\mu_{c}$, the onset of diquark condensation, at which point the system develops a Bose-Einstein condensation (BEC) phase.  As the chemical potential increases beyond $\mu \simeq m_{\pi}$, the dynamical mass of quark quasiparticles drops below $\mu$ and the system undergoes a BEC-BCS crossover.  With $\mu$ further increasing, the quark quasiparticle energy develops a minimum at finite momentum $p$.

Above the deconfinement transition, the EOS is governed by quark degrees of freedom, but the hadronic modes still have a significant influence up to the crossover temperature range.

The quark density $\rho$ as function of the chemical potential is a quantity of interest that is readily accessible in two-color lattice QCD.  It turns out that the mean field approximation of the (P)NJL model underestimates this density by about a factor of two.  Corrections treated in Gaussian approximation do not change this result significantly.  This is not surprising since the Gaussian approximation misses important correlations between diquarks and scalar bosons generated by the higher order terms ($\mcl{L}^{(3)}$ and $\mcl{L}^{(4)}$) of the hadronic effective Lagrangian.  The net attractive interactions produced by strong couplings in $\mcl{L}^{(3)}$ and  $\mcl{L}^{(4)}$ are expected to raise the density considerably.  Work along these lines using Gaussian functional methods is in progress and will be reported elsewhere.  Nonetheless, the present work establishes a baseline for the further systematic expansion of fluctuations and correlations beyond mean field.

\vspace{1cm}
\section*{Acknowledgement}
S.I. and H.T. are grateful to Wolfram Weise and the colleagues of the theory group at Technische Universit\"{a}t M\"{u}nchen for fruitful discussions and warm hospitality extended to both of us.  S.I. acknowledges the support of Osaka University and H.T. of the Humboldt foundation.  One of us (W.W.) thanks Simon Hands for valuable information.  This work is partly supported by JSPS (21540267) and by the DFG cluster of excellence ''Origin and Structure of the Universe''.

\appendix

\section{Meson and diquark-baryon fields and propagators}

Generic terms emerging in the expansion of the logarithmic term $\tr\ln(1+\hat{S}\hat{K})=\sum_{k}U^{(k)}$ are given in this Appendix.   We write here the $k=2$ term for the kinetic and mass terms, the $k=3$ term for the coupling terms and the $k=4$ term for the interaction terms. 

\subsection{Pion}

The result of the Fourier transformed propagator for the pion in Eq. \eqref{k2} is:
\beq
\Gamma_{pp}(q)&=&\frac{i}{4}\tr\int\f{p}{4}g_0^2(H_{p^\prime}^-\gamma_5H_{p}^+\gamma_5+H_{p^\prime}^+\gamma_5H_{p}^-\gamma_5 \nn&& -G_{p^\prime}^+\gamma_5G_{p}^+\gamma_5
-G_{p^\prime}^-\gamma_5G_{p}^-\gamma_5)~,
\eeq
where $p'$ denotes $p+q$ with $q$ the outer four momentum.  
We have to take care of the spinor trace and the divergent integrals $I_2$ and $I_0$.  For $q=0$ one finds:
\beq
\Gamma_{pp}(0)&=&
i g_0^2 \tr_{fc} \int \f{p}{4} \frac{(p_0+E_p^-)(p_0-E_p^+)}{(p_0^2-(E_\Delta^{-})^{2})(p_0^2-(E_\Delta^{+})^{2})}\nn&&+\frac{(p_0-E_p^-)(p_0+E_p^+)-2\Delta^2}{(p_0^2-(E_\Delta^{-})^{2})(p_0^2-(E_\Delta^{+})^{2})}\nn
&=&g_0^2I_2-2g_0^2\mu^2I_0~.
\eeq
The first order derivative term is zero and we work out the second order derivative term.  After a lengthy manipulation we get
\beq
 \partial_\mu\partial_\nu\Gamma_{pp}(q)|_{q=0}&=g_0^2I_0 g_{\mu\nu}~.
\eeq

\subsection{Sigma boson}
Next we consider the $\sigma$ boson term:
\beq
 \Gamma_{ss}(q)=\frac{i}{4} \tr \int \f{p}{4} g_0^2 (G_{p^\prime}^+G_{p}^++G_{p^\prime}^-G_{p}^- + H_{p^\prime}^- H_{p}^+ +H_{p^\prime}^+H_{p}^-)~.
\eeq
In the $q=0$ limit: 
\beq
\Gamma_{ss}(0)=g_0^2I_2-2g_0^2\mu^2I_0-2g_0^2m^2I_0~.
\eeq
The first order derivative term vanishes.  
The result of the second order derivative of $\Gamma_{ss}$ is
\beq
 \partial_\mu\partial_\nu\Gamma_{ss}(q)|_{q=0}=g_0^2I_0 g_{\mu\nu}~.
\eeq
The pion field does not mix with other fields due to its pseudoscalar nature.  However, the sigma boson field mixes with diquark and antidiquark fields.  The mixing terms will be calculated later.

\subsection{Diquark-baryon}

We calculate the $d d^\ast $ term
\beq
 \Gamma_{dd^\ast}(q)&=&\frac{i}{4}\tr\int\f{p}{4}g_d^2(-G_{p^\prime}^+\gamma_5t_2\tau_2G_{p}^-\gamma_5t_2\tau_2)~.
\eeq
For $q=0$, we get
\beq
\Gamma_{dd^\ast}(0)&=&\frac{i}{2}g_d^2\tr\int\f{p}{4}\left(\frac{p_0^2-(E_p^-)^2}{(p_0^2-(E_{\Delta}^{-})^{2})^2}+\frac{p_0^2-(E_p^+)^2}{(p_0^2-(E_{\Delta}^{+})^{2})^2}\right)\nn
&=&\frac{1}{2}g_d^2I_2-\frac{1}{2}g_d^4|\Delta_{0}|^2I_0~.
\eeq
The first order time derivative is
\beq
\partial_0\Gamma_{dd^\ast}|_{q=0}&=&\frac{i}{2}g_d^2\tr\int\f{p}{4}\left[\left(\frac{E_p}{(p_0^2-(E_{\Delta}^{+})^{2})^2}-\frac{E_p}{(p_0^2-(E_{\Delta}^{-})^{2})^2}\right)\right.
\nn&&\left.+\left(\frac{\mu}{(p_0^2-(E_\Delta^{+})^{2})^2}+\frac{\mu}{(p_0^2-(E_\Delta^{-})^{2})^2}\right)\right]\nn&=&
g_d^2\mu I_0~.
\eeq
The second order derivative of $\Gamma_{dd^\ast}$ is
\beq
 \partial_\mu\partial_\nu\Gamma_{dd^\ast}(q)|_{q=0}=\frac{1}{2}g_d^2I_0g_{\mu\nu}~.
\eeq
Next, the $d^\ast d$ term is:
\beq
\Gamma_{d^\ast d}(q)=\frac{i}{4}g_d^2\tr\int\f{p}{4}(-G_{p^\prime}^-\gamma_5t_2\tau_2G_p^+\gamma_5t_2\tau_2)~.
\eeq
For $q=0$, we find
\beq
\Gamma_{d^\ast d}(0)=\frac{1}{2}g_d^2I_2-\frac{1}{2}g_d^4|\Delta_{0}|^2I_0~.
\eeq
The first order time derivative becomes:
\beq
\partial_0\Gamma_{d^\ast d}(q)|_{q=0}&=&-g_d^2\mu I_0~.
\eeq
The second order derivative term is:
\beq
 \partial_\mu\partial_\nu\Gamma_{d^\ast d}(q)|_{q=0}=\frac{1}{2}g_d^2I_0g_{\mu\nu}~.
\eeq
Now we consider the $dd$ and $d^\ast d^\ast$ terms that describe diquark mixing.  The $dd$ term is:
\beq
\Gamma_{dd}(q)=\frac{i}{4}\tr\int\f{p}{4}g_d^2H_{p^\prime}^+\gamma_5t_2\tau_2H_{p}^+\gamma_5t_2\tau_2~.
\eeq
For $q=0$, we get
\beq
\nonumber
\Gamma_{dd}(0)&=&\frac{i}{2}g_d^4(\Delta_0^\ast)^2\tr\int\f{p}{4}\left(\frac{1}{(p_0^2-(E_{\Delta}^{+})^{2})^2}+\frac{1}{(p_0^2-(E_{\Delta}^{-})^{2})^2}\right)\\
&=&-\frac{1}{2}g_d^4(\Delta_0^\ast)^2I_0~.
\eeq
All the derivative terms are zero.  By analogy the $d^*d^*$ term becomes: 
\beq
\Gamma_{d^\ast d^\ast}(q)&=&\frac{i}{4}\tr\int\f{p}{4}g_d^2H_{p^\prime}^-\gamma_5t_2\tau_2H_{p}^-\gamma_5t_2\tau_2~,\nn
\Gamma_{d^\ast d^\ast}(0)&=&-\frac{1}{2}g_d^4(\Delta_0)^2I_0~.
\eeq

\subsection{Sigma-diquark mixing}

The pion does not mix with diquark-baryon fields due to its pseudo-scalar nature.  However, the sigma boson mixes with diquark fields when diquark condensation takes place.  We calculate the mixing terms $sd, ds, sd^\ast$ and $d^\ast s$.
\beq
\Gamma_{ds}(q)&=&\frac{i}{4}\tr\int\f{p}{4}(-g_0g_d)(G_{p^\prime}^+\gamma_5t_2\tau_2H_p^++H_{p^\prime}^+\gamma_5t_2\tau_2G_{p}^-)~.
\eeq
For $q=0$, we get
\beq
\Gamma_{ds}(0)=-g_0g_d^2\Delta_0^\ast mI_0~.
\eeq
For the $sd$ term,
\beq 
 \Gamma_{sd}(q)=\frac{i}{4}\tr\int\f{p}{4}(-g_0g_d)(H_{p^\prime}^+G_p^+\gamma_5t_2\tau_2+G_{p^\prime}^-H_{p}^+\gamma_5t_2\tau_2)~.
\eeq
At $q=0$:
\beq
\Gamma_{sd}(0)=-g_0g_d^2\Delta_0^\ast m I_0~.
\eeq
The derivatives of $\Gamma_{ds}$ and $\Gamma_{sd}$ terms are finite.  We calculate the $d^\ast s$ term
\beq
\Gamma_{d^\ast s}(q)=\frac{i}{4}\tr\int\f{p}{4}g_0g_d(G_{p^\prime}^-\gamma_5t_2\tau_2H_p^-+H_{p^\prime}^-\gamma_5t_2\tau_2G_p^+)~.
\eeq
For $q=0$, the result is:
\beq
\Gamma_{d^\ast s}(0)=-g_0g_d^2\Delta_0 m I_0~.
\eeq
Next, the $sd^\ast$ term:
\beq
\Gamma_{sd^\ast}(q)=\frac{i}{4}\tr\int\f{p}{4}g_0g_d(H_{p^\prime}^-G_p^-\gamma_5t_2\tau_2+G_{p^\prime}^+H_{p}^-\gamma_5t_2\tau_2)~.
\eeq
Its $q=0$ limit is:
\beq
 \Gamma_{sd^\ast}(0)=-g_0g_d^2\Delta_0 m I_0~.
\eeq
All $k=2$ terms are collected in Eq. (\ref{second}).

\subsection{The third order term $U^{(3)}$}

Here we examine $U^{(3)}=-\frac{i}{6}\tr(\hat{S}\hat{K})$. This term generates leading combinations of $G^{\pm}$ and $H^{\pm}$ with $\gamma_5$.  With the principle of taking only the divergent integrals, the possible combinations are $GGG$ or $GGH$ with $\gamma_5$.  From dimensional analysis, the coupling constant should have mass dimension one.  Thus for baryon-number conserving channels, the relevant factor is $m$.   For baryon-number non-conserving channels the factors are $\Delta_0$ or $\Delta_0^*$. Hence, the terms for baryon-number conserving channels are generated from $GGG$ and the terms for baryon-number non-conserving channels are generated from $GGH$. We find 32 terms that remain finite from the above discussion. We do not write all the terms, but give a representative example for a baryon-number conserving channel:
\beq
U^{(3)}_{s^3}&=&\frac{i}{6}g_0^3\tr\int\f{p}{4}(G^{+}G^{+}G^{+}+G^{-}G^{-}G^{-})s^3\nn
&=&-2g_0^3I_0ms^3~,
\eeq
and for a baryon-number non-conserving channel:
\beq
U^{(3)}_{d^2d^*}&=&\frac{i}{6}g_d^3\tr\int\f{p}{4}(G^{+}\gamma_5H^{'+}\gamma_5G^{-}\gamma_5+H^{'+}\gamma_5G^{-}\gamma_5G^{+}\gamma_5
\nn
&&+G^{-}\gamma_5G^{+}\gamma_5H^{'+}\gamma_5)d^2d^*
=-g_d^3I_0\Delta^*_0d^2d^* ~,
\eeq
where
\beq
H^{'+}=-\frac{ g_d\Delta_0^{*}}{p_0^2-(E_\Delta^{\pm})^{2}}\gamma_5\tilde{\Lambda}_+
-\frac{ g_d\Delta_0^{*}}{p_0^2-(E_\Delta^{\mp})^{2}}\gamma_5\tilde{\Lambda}_-~.
\eeq
The final result for $k=3$ is displayed in Eq. \eqref{third}.

\subsection{The fourth order term $U^{(4)}$}

Next, consider $U^{(4)}=\frac{i}{8}\tr(\hat{S}\hat{K})^4$.   By examination it turns out that  the only possible combination is $GGGG$ and the coupling constants are dimensionless.  Out of the many terms that appear we give here a generic example:
\beq
U^{(4)}_{\pi}&=&\frac{i}{8}g_0^4\tr\int\f{p}{4}(G^{+}\gamma_5G^{+}\gamma_5G^{+}\gamma_5G^{+}\gamma_5+ (+\rightarrow -))\vec{\pi}^4~.
\label{k4pi}
\eeq
The products of gamma matrices are worked out as follows:
\beq
\Lambda_-\gamma_0\gamma_5\Lambda_+\gamma_0\gamma_5\Lambda_-\gamma_0\gamma_5\Lambda_+\gamma_0\gamma_5&=&\Lambda_- ~,\nn
\Lambda_+\gamma_0\gamma_5\Lambda_-\gamma_0\gamma_5\Lambda_+\gamma_0\gamma_5\Lambda_-\gamma_0\gamma_5&=&\Lambda_+ ~,
\eeq
and all other terms vanish.
From the first term one finds:
\beq
\tr_{s}G^{+}\gamma_5G^{+}\gamma_5G^{+}\gamma_5G^{+}\gamma_5=2\left(\frac{1}{(p_0^2-(E_{\Delta}^{-})^{2})^2}+\frac{1}{(p_0^2-(E_{\Delta}^{+})^{2})^2}\right) ~.
\eeq
The factor 2 in this equation comes from the spinor trace.  The same result is obtained from the second term of Eq. \eqref{k4pi}, hence the result of the complete $\vec{\pi}^4$ term is
\beq
&& \frac{i}{2} g_0^4\tr_{fc} \int\f{p}{4}\left(\frac{1}{(p_0^2-(E_{\Delta}^{-})^{2})^2}+\frac{1}{(p_0^2-(E_{\Delta}^{+})^{2})^2}\right)\vec{\pi}^4\nn&&=-\frac{1}{2}g_0^4I_0\vec{\pi}^4 ~.
\eeq
Consider another example, the combination involving  $\vec{\pi}^2dd^*$, which has 12 terms of the form $GGGG$. By repeating the above procedure, one finds:
\beq
&& i g_0^2g_d^2\tr_{fc}\int \f{p}{4} \left(\frac{1}{(p_0^2-(E_{\Delta}^{-})^{2})^2}+\frac{1}{(p_0^2-(E_{\Delta}^{+})^{2})^2}\right)\vec{\pi}^{2}dd^*\nn&&=-g_0^2g_d^2I_0\vec{\pi}^{2}dd^*~.
\eeq
The complete result for $k=4$ is written in Eq. \eqref{fourth}.

\section{Hadron masses}
\label{app2}

Here we derive the mass spectrum of hadrons.  The pion does not mix with other hadrons, but the sigma boson mixes with diquark-baryons once diquark condensation takes place.  We list all the masses of mesons and diquark baryons.

\subsection{Pion and sigma masses}\label{pisigma}

The renormalized pion mass is written as 
\beq
m_{\pi}^2&=(M_{s}^2-2g_0^2I_2+4g_0^2I_0\mu^2)/(g_{0}^{2}I_0)~.
\eeq
From hereon, all the masses are understood to be renormalized ones.   Consider the case of $|\Delta_0|=0$ and define divergent integrals for $\mu=0$ as:
\beq
 I_{2}^0=2i\tr_{fc}\int\f{p}{4}\frac{1}{p_0^2-E_p^2}\quad I_{0}^0=-2i\tr_{fc}\int\f{p}{4}\frac{1}{(p_0^2-E_p^2)^2}~.
\eeq
$I_2$ for finite $\mu$ with $|\Delta_0|=0$ is written in terms of $I_{2}^0$ and $I_{0}^0$
\beq
I_2&=& i \tr_{fc}\int \f{p}{4}\left(\frac{1}{p_0^2-(E_p^{+})^{2}}+\frac{1}{p_0^2-(E_p^{-})^{2}}\right)\nn
&=& I_{2}^0+2\mu^2I_{0}^0~.
\eeq
In the region of interest, $I_0=I_{0}^{0}$ and $m_{\pi}^2$ becomes
\beq
m_{\pi}^2=(M_{s}^{2}-2 g_{0}^{2}I_{2}^0) /(g_{0}^{2}I_{0}^0) \equiv m_{\pi0}^2~.
\eeq
Hence, the pion mass does not depend on $\mu$ for $|\Delta_0|=0$.

The diquark gap equation with the Pauli-G\"{u}rsey symmetry $G_0=H_0$ provides $g_{0}=g_{d}$ and $M_{s}=M_{d}$ and therefore $M_{s}^2-2g_0^2I_2=0$ for the finite $|\Delta_0|$ in which $m_{\pi}^2=4\mu^2$.  This means that the pion mass in the diquark condensed phase increases linearly with $\mu$.

As for the sigma mass, $m_s^2$ is larger than $m_{\pi}^{2}$ by twice dynamical quark mass squared, $4m^2$, for the case without sigma-diquark mixing:
\beq
m_s^2=(M_{s}^2-2g_0^2I_2+4g_0^2I_0\mu^2+4g_0^2I_0m^2)/(g_{0}^{2}I_{0})=m_{\pi}^2+4m^2~.
\eeq
Hence as the dynamical quark mass approaches zero, these two masses coincide.  The sigma boson, however, mixes with diquark-baryons once diquark condensation sets in.

\subsection{Diagonalization of sigma and diquark-baryon mass matrix}

The sigma boson mixes with diquark-baryons, and furthermore diquarks mix with each other.  The sigma and diquark Lagrangian density with kinetic and mass terms is written in matrix form:
\beq
\nonumber
\mcl{L}_{sd}^M&=& -\frac{1}{2}
\Phi^{\dagger}
\begin{pmatrix}
  \partial^2+m_s^2 &2\Delta^* m&2\Delta m\\
 2\Delta m&\frac{1}{2}(\partial^2+m_d^2)+2i\mu\partial^0&\Delta^2\\
 2\Delta^* m&\Delta^{*2}&\frac{1}{2}(\partial^2+m_d^2)-2 i\mu \partial^0
\end{pmatrix}
\Phi~,\\
&&
\eeq
with the scalar fields representation $\Phi^t=(s, d, d^*)$.
The diquark-baryon mass becomes:
\beq
m_{d}^{2}&=&(M_{d}^2-2g_d^2I_2+2g_d^2I_0|\Delta|^2)/(g_{0}^{2}I_0)\nn
&=&
\begin{cases}
 m_{\pi}^{2}-4\mu^2 & |\Delta|=0~,\\
 2 |\Delta|^2 & |\Delta|\neq 0~,
\end{cases}
\eeq
using the diquark gap equation $M_d^2-2g_d^{2}I_2=0$ and $g_d=g_0$ for $|\Delta|\neq 0$.

We write the mass matrix in the momentum representation (Minkowski space) as
\beq
\label{matrix2}
 M(\omega,\vec{q})=
\begin{pmatrix}
  {q}^2-m_s^2&-2\Delta^* m& -2\Delta m\\
 -2\Delta m&\fra({q}^2-m_d^2)+2\mu\omega&-\Delta^2\\
 -2\Delta^* m&-\Delta^{*2}&\fra({q}^2-m_d^2)-2\mu\omega
\end{pmatrix}~,
\eeq
where $q^{2}=\omega^{2}-\vec q^{\,2}$.
The mass spectra for sigma, diquark and antidiquark are obtained solving the dispersion relation $\det M(\omega,0)=0$ with respect to $\omega^{2}$.
One of the solutions is zero and identified with the Nambu-Goldstone (NG) boson at $|\Delta|\neq0$, since the baryon number symmetry is spontaneously broken.
The other solutions are given analytically as
\beq
\label{omega1}
\omega^2=2 |\Delta|^2+10\mu^2+2m^2\pm\sqrt{(6\mu^2+2 |\Delta|^2+2m^2)^2-48\mu^2m^2}~.
\eeq
In the limit of $|\Delta|\rightarrow 0$ at the onset of diquark condensation $(\mu=m_{\pi}/2)$ the dispersion Eq. \eqref{omega1} becomes 
\beq
\omega^2=
\begin{cases}
 4\mu^2+4m^2=m_s^2~.\\
 16\mu^2=4m_{\pi}^2~.
\end{cases}
\eeq
The upper solution is the squared sigma mass $m_s^2$ (without mixing effect) and the lower one is the squared antidiquark mass at $\mu=m_{\pi}/2$.   In the limit  $m\rightarrow 0$ at large $\mu$ the dispersion Eq. \eqref{omega1} gives
\beq
\omega^2=
\begin{cases}
 4\mu^2=m_{\pi}^2~.\\
 16\mu^2+4|\Delta|^2~.
\end{cases}
\eeq
One solution represents the linearly increasing pion mass.  The other one is the squared antidiquark baryon mass at large $\mu$ where mixing effects with the scalar boson disappears.

When $\Delta=0$ $(\mu \leq m_{\pi}/2)$ the mixing terms are zero.  
The diquark masses are obtained as
\beq
\omega&=&m_{\pi}\pm 2\mu~.
\label{diq}
\eeq
The diquark and antidiquark masses move linearly with $\mu$ for $\Delta=0$. The diquark-baryon behaves as $m_{\pi}-2\mu$ and becomes the NG boson at non-zero $|\Delta|$.  On the other hand, the antidiquark with baryon number $-1$ has the mass $m_{\pi}+2\mu$.
These mass spectra including the pion are plotted in Fig. \ref{hadronmass}.

In the absence of the sigma-diquark mixing term, $-2m\Delta^{(*)}$, the diquark masses become
\beq
\label{diqm}
\omega_{\Delta^*}&=&\sqrt{16\mu^2+4|\Delta|^2}~,\nn
\omega_{\Delta}&=&0~.
\eeq
The sigma mass approaches the pion mass as the dynamical quark mass tends to zero due to the chiral symmetry restoration.
This feature is shown in Fig. \ref{hadronmass_wom}.

\subsection{The Bogoliubov mode}

The hadronic contributions to the pressure and the quark density involve spectra of hadrons.  It is important to know how the zero mode behaves at finite momentum in the diquark-condensed phase.  The Bogoliubov excitation~\cite{Brauner2008,He2010,Diener2008} for the zero mode results from the solution $\omega(\vec q\,)$ of $\det M(\omega,\vec{q})=0$ in Eq. \eqref{matrix2} at finite momentum $\vec{q}$ as
\beq
\det M(\omega,\vec{q})&=&
\omega^6-[(\vec{q}\,^2+m_s^2)+2(\vec{q}\,^2+m_d^2+8\mu^2)]\omega^4\nn
&&+[\vec{q}\,^2(\vec{q}\,^2+2m_d^2)+2(\vec{q}\,^2+m_s^2)(\vec{q}\,^2+m_d^2+8\mu^2)-16m^2|\Delta|^2]\omega^2\nn
&&-\vec{q}\,^2[(\vec{q}\,^2+m_s^2)(\vec{q}\,^2+2m_d^2)+16m^2|\Delta|^2]=0~.
\eeq
We write the solutions as $\omega_i$ $(i=s,\,d,\,d^{*})$.
Dropping the sigma-diquark mixing terms in the full mass matrix (\ref{matrix2}), one arrives at an analytical expression for $\omega(\vec q\,)$ of the Bogoliubov mode.  Consider the reduced mass matrix
\beq
D(\omega,\vec{q})=
\begin{pmatrix}
 \omega^2-\vec{q}\,^2-m_d^2+4\mu\omega & -2\Delta^{2}\\
 -2\Delta^{*2}&\omega^2-\vec{q}\,^2-m_d^2-4\mu\omega
\end{pmatrix} ~,
\eeq
and solve the dispersion relation $\det D(\omega,\vec{q}\,)=0$ with respect to $\omega(\vec q\,)$ for the lowest mode:
\beq
\omega^2&=&\vec{q}\,^2+2|\Delta|^2+8\mu^2\nn && -\sqrt{(\vec{q}\,^2+2|\Delta|^2+8\mu^2)^2 -\vec{q}\,^2(\vec{q}\,^2+4|\Delta|^2)}\nn
&\sim&\frac{\vec{q}\,^2(\vec{q}\,^2+4|\Delta|^2)}{2(\vec{q}\,^2+2|\Delta|^2+8\mu^2)}~.
\eeq
In the BEC limit, the pion mass $m_{\pi}=2\mu$ is much larger than $|\Delta|$. The Bogoliubov excitation is then written as
\beq
\label{bog}
\omega(\vec{q}\,)\sim\sqrt{\frac{\vec{q}\,^2}{2m_{\pi}}\left(\frac{\vec{q}\,^2}{2m_{\pi}}+\frac{2|\Delta|^2}{m_{\pi}}\right)}~.
\eeq
The zero mode varies linearly with $|\vec{q}\,|$ at small momentum.  With inclusion of the sigma-diquark coupling, the dispersion equation for $\omega(\vec q\,)$ with the full mass matrix is solved numerically.  For $m_{\pi}\gg |\Delta|$ one confirms that the zero mode has the Bogoliubov excitation spectrum as expressed in Eq. (\ref{bog}).


\end{document}